\RequirePackage[2020-02-02]{latexrelease}
\documentclass[aps,11pt,preprintnumbers,nofootinbib,floatfix]{revtex4}

\usepackage{subfigure}
\usepackage{multirow}
\usepackage{amsmath}

\usepackage{amssymb}
\usepackage{graphicx}
\begin{document}

\title{Generalized free energy and thermodynamic phases of black holes in the gauged Kaluza-Klein theory}

%%%% To generate auto affiliation numbers please use \author{}\affil{} command
\author{Tran N. Hung}
\email{hung.tranngoc@phenikaa-uni.edu.vn}  
\affiliation{Phenikaa Institute for Advanced Study and Faculty of Fundamental Sciences, Phenikaa University, Yen Nghia, Ha Dong, Hanoi 12116, Vietnam}
\author{Cao H. Nam}
\email{nam.caohoang@phenikaa-uni.edu.vn}  
\affiliation{Phenikaa Institute for Advanced Study and Faculty of Fundamental Sciences, Phenikaa University, Yen Nghia, Ha Dong, Hanoi 12116, Vietnam}
\date{\today}

\begin{abstract}
In the context of the generalized (off-shell) free energy, we explore the phase emergence and corresponding phase transitions of charged dilaton $\text{AdS}$ black holes in the gauged Kaluza-Klein (KK) theory where the KK vector field is gauged such that the fermionic fields are charged under the U(1)$_{\text{KK}}$ gauge group. The black hole solutions are asymptotic to the AdS$_D$ geometry and can be realized as the dimensional reduction of the gauged supergravities on the compact internal manifolds, leading to the restriction as $4\leq D\leq 7$. By studying the behavior of the generalized free energy under the change of the ensemble temperature, we determine the thermodynamic phases and the corresponding phase transitions of black holes. This is confirmed by investigating the heat capacity at the constant pressure and the on-shell free energy. In the canonical ensemble, the thermodynamics of black holes can be classified into three different classes as follows: (i) $D=4$, (ii) $D=5$, and (iii) $D=6,7$. Whereas, in the grand canonical ensemble, the thermodynamics of black holes is independent of the number of spacetime dimensions and the pressure, but depends on the chemical potential $\Phi$. The thermodynamic behavior of black holes can be classified into three different classes as follows: (i) $\Phi<1$, (ii) $\Phi>1$, and (iii) $\Phi=1$.

\end{abstract}

\maketitle

\section{Introduction}

The Kaluza-Klein (KK) theory provides a beautiful and elegant way to unify the gravitational interaction with the non-the gravitational interactions mediated by the vector fields in a geometric framework \cite{Witten1981,Bailin1987,Overduin1997}. In this context, the gauge symmetry of the non-gravitational interactions is the isometry of the compact internal space where the number of the generators of the gauge symmetry is equal to the number of the Killing vectors of the internal manifold. The vector fields mediating the non-gravitational interactions describe the excitations of the off-diagonal components of the metric of the higher-dimensional spacetime. In the phenomenological aspect, the KK gauge bosons can lead to the emergence of new neutral gauge bosons beyond the Standard Model which are being searched for at the colliders \cite{Nam2019,Nam2020} as well as the candidate for the dark matter \cite{Nam2023}. Whereas, the radion (or dilaton) fields can be responsible for the inflaton \cite{Cline2000,Kolb2003,Sundrum2010,Trudeau2012,Fukazawa2013,Nam2021}. In addition, the KK reduction on a circle can produce the black hole solutions without the unphysical curvature singularity \cite{Bah2021a,Bah2021b} and the microstates for the Bekenstein-Hawking entropy of the observed black holes \cite{Nam2023b}.

The KK theory which is the dimensional reduction of the pure gravity on the circle is a particularly interesting class of the Einstein-Maxwell-dilaton (EMD) theories whose black hole solutions can be embedded in string theory/supergravity \cite{Horo1991,Dgar1991,Behr1999,Duff199,Lu1999,Cvetic1999}. Interestingly, the KK vector field can be gauged such that the fermion fields carry the U(1)$_{\text{KK}}$ charge, resulting in the presence of the Killing spinor equations in the KK theory \cite{Pope2011}. This gauge process allows us to embed the KK theory in gauged supergravities in some dimensions. In addition, gauging the KK vector field generates a scalar potential for the dilaton field which is restricted to exponential terms by the projected integrability condition of the Killing spinor equations. The scalar potential which is an extension of the cosmological constant plays a pivotal role in changing the asymptotic behavior of the black hole solutions. The black hole solutions in the gauged KK theory are asymptotic to the anti-de Sitter (AdS) geometry where the curvature radius of asymptotic AdS space is determined in terms of the KK gauge coupling \cite{Liu2012}. 

It was shown that new thermodynamic aspects and phase transitions of black holes emerge as the cosmological constant is realized as the thermodynamic pressure leading to the extended phase space \cite{MTeo2017}. In this context, the ADM mass of black holes is identified as the enthalpy rather than the internal energy. New thermodynamic phenomena of black holes emerge in the extended phase space such as the $P-V$ criticality \cite{Mann2012,Cai2013,Hendi2013,Fernando2016,Hendi2017,CNam2019}, the reentrant phase transition \cite{Mann2013}, the triple point \cite{Altamirano2014,WeiLiu2014}, the black hole heat engine \cite{Johson2014,Bhamidipati2017,YLi2018,QLan2018}, and the Joule-Thomson expansion \cite{Aydner2017,Aydner2018,XMo2018,CNam2020}. Inspired by the AdS/CFT correspondence \cite{Maldacena1998}, the thermodynamics and phase transitions of AdS black holes have been extensively investigated to understand the universal properties of the strongly interacting systems on the AdS boundary \cite{Ahmed2023,WCong2023}.

Recently, the generalized free energy extended from the on-shell free energy has been proposed for study the thermodynamic phases and phase transitions of black holes \cite{JWang2020,KZhang2020,KZhang2021,WeiLiu2022,LiWang2022,CLiu2023,MSAli2023,LiWang2023,Ksafir2023,WangXu2024}. In this formalism, the generalized free energy depends on the event horizon radius which is treated as an order parameter which is analogous to the density playing the role of the older parameter in the liquid-gas system. In this way, the ensemble at a specific temperature comprises all black hole states with various values of event horizon radius from zero to infinity. Note that, the black hole state with the zero event horizon radius is the thermal AdS space. By investigating the behavior of the generalized free energy in the event horizon radius when changing the ensemble temperature, we can determine the phase emergence of the black holes, their thermodynamic stability, and the corresponding phase transitions. The extremal points of the generalized free energy represent the on-shell black hole states which are the solutions of equations of motion. The local maximum points correspond to the thermodynamically unstable black holes which have a negative heat capacity. The local and global minimum points represent black holes corresponding to the local and global stationary states with a positive heat capacity, respectively. The thermal fluctuations in the event horizon radius can be generated in a stochastic way in which the force driving the stochastic process is encoded in the generalized free energy. As a result, it is necessary to include the off-shell black hole states which are not the solutions of equations of motion. The presence of the fluctuating black holes hence leads to the phase transitions.

In the present paper, we explore the emergence of the thermodynamic phases and the corresponding phase transitions of charged dilaton $\text{AdS}$ black hole in the gauged KK theory by using the formalism of the generalized free energy. In the canonical ensemble where the black hole charge is kept fixed, the thermodynamics and phase transitions of black holes depend on the number of spacetime dimensions. In this situation, we classify the thermodynamic behavior of black holes into three different classes corresponding to $D=4$, $D=5$, and $D=6,7$. In the grand canonical ensemble where the chemical potential is kept fixed, we show that the thermodynamics and phase transitions of black holes depend only on the chemical potential $\Phi$ and are classified in the classes as follows $\Phi<1$, $\Phi>1$, and $\Phi=1$. In addition, we study the behavior of the heat capacity and the on-shell free energy to confirm the thermodynamic stability and the phase transitions that are observed from the behavior of the generalized free energy as changing the ensemble temperature.

This paper is organized as follows. In Sec. \ref{sec:solution}, we review charged dilaton $\text{AdS}$ black holes in the gauged KK theory, and then we represent their thermodynamic rules in the extended phase space. In Sec. \ref{sec:CE}, we study the thermodynamics of black holes in the canonical ensemble where we classify the emergence of the thermodynamic phases and the corresponding phase transitions based on the number of spacetime dimensions. In Sec. \ref{sec:GCE}, we represent the thermodynamics of black holes in the grand canonical ensemble where the classification for the phase emergence and phase transitions is based on the chemical potential. Finally, we make conclusions in Sec. \ref{sec:conclu}.

\section{Charged dilaton $\text{AdS}$ black holes in the gauged KK theory}
\label{sec:solution}
We consider the $D$-dimensional effective field theory derived from the KK reduction of $(D+1)$-dimensional pure Einstein gravity on a circle $S^1$. The corresponding action is given by
\begin{eqnarray}
    S_D=\int d^Dx\sqrt{-g}\left[\mathcal{R}-\frac{1}{2}g^{\mu\nu}\partial_\mu\phi\partial_\nu\phi-\frac{1}{4}e^{a\phi}F_{\mu\nu}F^{\mu\nu}-V(\phi)\right],\label{KKLagra}
\end{eqnarray}
where $g$ denotes the determinant of the $D$-dimensional metric, $\mathcal{R}$ is the Ricci scalar, $\phi$ is the dilaton, $F_{\mu\nu}=\partial_\mu A_\nu-\partial_\nu A_\mu$ is the field strength tensor of the KK gauge field $A_\mu$, $a=\sqrt{2(D-1)/(D-2)}$ is the coupling between the dilaton and the KK gauge field, and $V$ is the potential of the dilaton. In the usual KK reduction, the potential of the dilaton is zero, i.e. $V(\phi)=0$. The equations of motion for the metric $g_{\mu\nu}$, the KK gauge field $A_\mu$, and the dilaton $\phi$ which are derived from the variation of the action (\ref{KKLagra}) read
\begin{eqnarray}
R_{\mu\nu}&=&\frac{1}{2}\partial_\mu\phi\partial_\nu\phi-\frac{V(\phi)}{D-2}g_{\mu\nu}-\frac{e^{a\phi}}{2}\left(F_{\mu\rho}{F_\nu}^\rho-\frac{F_{\rho\lambda}F^{\rho\lambda}}{2(D-2)g_{\mu\nu}}\right),\label{Eins-eqs}\\
0&=&\partial_\mu\left(\sqrt{-g}e^{a\phi}F^{\mu\nu}\right),\label{vect-eqs}\\
\nabla^2\phi&=&V'(\phi)-\frac{e^{a\phi}}{4}F_{\mu\nu}F^{\mu\nu}.\label{scal-eqs}
\end{eqnarray}

It is possible to gauge the KK gauge field $A_\mu$ such that the fermionic fields are charged under the U(1)$_{\text{KK}}$ gauge group. This leads to a pseudo-supersymmetrization of the $D$-dimensional KK theory where the pseudo-gravitino and pseudo-dilatino are charged under U(1)$_{\text{KK}}$. Interestingly, gauging the KK gauge field allows us to generate a potential for the dilaton, in analogy to the gauged supergravity \cite{Liu2011}. In the bosonic background given by (\ref{KKLagra}), the pseudo-supersymmetric variation of the fermionic partners is zero, leading to the following Killing spinor equations \cite{Liu2012}
\begin{eqnarray}
D_\mu\eta+\frac{W(\phi)}{2\sqrt{2}(D-2)}\Gamma_\mu\eta+\frac{ie^{a\phi/2}}{8(D-2)}\left[\Gamma_\mu\Gamma^{\nu\rho}-2(D-2)\delta^\nu_\mu\Gamma^\rho\right]F_{\nu\rho}\eta\equiv\hat{D}_\mu\eta&=&0,\label{Killspeq1}\\
\left[\Gamma^\mu\partial_\mu\phi+\frac{i}{4}ae^{a\phi/2}\Gamma^{\mu\nu}F_{\mu\nu}-\sqrt{2}W'(\phi)\right]\eta&=&0,\label{Killspeq2}
\end{eqnarray}
where $D_\mu$ is the covariant derivative given by
\begin{eqnarray}
 D_\mu=\partial_\mu+\frac{i}{16}{\omega_\mu}^{\alpha\beta}[\Gamma_\alpha,\Gamma_\beta]-\frac{ig(D-3)}{4}A_\mu, 
\end{eqnarray}
with ${\omega_\mu}^{\alpha\beta}$ being the spin connection and $g$ to be the KK gauge coupling, the matrices $\Gamma_\mu$ are related to the $D$-dimensional gamma matrices $\Gamma_\alpha$ as $\Gamma_\mu={e^\alpha}_\mu\Gamma_\alpha$ with ${e^\alpha}_\mu$ are the vielbeins, $\Gamma_{\mu\nu}\equiv\frac{i}{4}[\Gamma_\mu,\Gamma_\nu]$, and $\eta$ is the Killing spinor. Eqs. (\ref{Killspeq1}) and (\ref{Killspeq2}) corresponds to the pseudo-supersymmetric transformation rules of the pseudo-gravitino and the pseudo-dilatino, respectively. Whereas, the pseudo-supersymmetric variation of the vielbeins, the KK gauge field, and the dilaton automatically vanish in the bosonic background. Note that, the second term on the left-hand side of Eq. (\ref{Killspeq1}) is inspired by supergravity with $W(\phi)$ referring to the superpotential. Accordingly, the dilaton $V(\phi)$ potential can be expressed in terms of the superpotential $W(\phi)$ as follows
\begin{eqnarray}
    V(\phi)=W'(\phi)^2-\frac{D-1}{2(D-2)}W(\phi)^2.\label{pot-suppot}
\end{eqnarray}

The superpotential $W(\phi)$ is fixed by the projected integrability condition by acting $\Gamma^\mu[\hat{D}_\mu,\hat{D}_\mu]$ on Eq. (\ref{Killspeq1}) and $\Gamma^\mu\nabla_\mu$ on Eq. (\ref{Killspeq2}). The projected integrability condition gives equations that consist of
the terms related to equations of motion (which automatically vanish) and the unrelated terms that have to vanish. Then, we determine the superpotential as follows
\begin{eqnarray}
    W(\phi)=\frac{(D-3)g}{\sqrt{2}}\left(e^{-a_1\phi}+\frac{a_1}{a_2}e^{a_2\phi}\right),
\end{eqnarray}
where $a_1=(D-1)/\sqrt{2(D-1)(D-2)}$ and $a_2=(D-3)/\sqrt{2(D-1)(D-2)}$. Using Eq. (\ref{pot-suppot}), we find the dilaton potential as
\begin{eqnarray}
  V(\phi)=-(D-1)g^2\left[(D-3)e^{-\frac{2a_1}{D-1}\phi}+e^{2a_2\phi}\right].\label{GKK-dilapot}
\end{eqnarray}
This potential possesses a maximal point at $\phi=0$ corresponding to a negative cosmological constant $\Lambda=-(D-1)(D-2)g^2/2$.

By solving Eqs. (\ref{Eins-eqs}), (\ref{vect-eqs}), and (\ref{scal-eqs}) with the dilaton potential (\ref{GKK-dilapot}), we can find a charged dilaton AdS black hole solution that is spherically symmetric and static in the gauged KK theory as follows \cite{Liu2012}
\begin{eqnarray}
    ds^2&=&-H(r)^{-\frac{D-3}{D-2}}f(r)dt^2 + H(r)^{\frac{1}{D-2}}\left(\frac{dr^2}{f(r)}+r^2 d\Omega_{D-2}^{2}\right),\nonumber\\
    A_\mu&=&\sqrt{1+\frac{m}{q}}\frac{1}{H(r)}\delta^t_\mu\equiv\phi(r)\delta^t_\mu,\label{gauKK-bhsol}\\
    \phi&=&\frac{a}{2}\log H(r),\nonumber
\end{eqnarray}
where
\begin{eqnarray}
    H(r)&=&1+\frac{q}{r^{D-3}},\label{Hr:exp}\\
    f(r)&=&1-\frac{m}{r^{D-3}}+g^2 r^2 H(r),\label{fr:exp}
\end{eqnarray}
$m$ and $q$ are the integral constants related to the ADM mass and charge of black holes, and $d\Omega_{D-2}^{2}$ is the line element of unit $(D-2)$-sphere. In the limit of $g\rightarrow0$ corresponding to the usual KK theory, the black hole solution given in (\ref{gauKK-bhsol}) becomes the non-rotating black hole solution that was found in \cite{Gibbons1987}. In addition, one can recover the Schwarzschild solution when the gauge coupling $g$ goes to zero and the black hole charge vanishes.

The black hole solution given by Eq. (\ref{gauKK-bhsol}) is asymptotic to the AdS$_D$ geometry. Interestingly, this black hole solution has a string/M theory origin as the dimensional reduction of string/M theory on the compact internal manifolds \cite{Tran1999}. The cases of $D=4$ and $D=7$ can be derived from the compactification of eleven-dimensional supergravity on $S^7$ and $S^4$, respectively. The $D=5$ case can come from the dimensional reduction of IIB supergravity on $S^5$. Whereas, the $D=6$ case can embed in the $F(4)$ gauged supergravity \cite{Romans1986}. Motivated by the string/M theory origin, in this work, we are only interested in studying the emergence of the thermodynamic phases and corresponding phase transitions of black holes for $4\leq D\leq 7$.

The horizon radius $r_h$ of black holes is given by the root of the following equation, $f(r_h)=0$, which leads to
\begin{eqnarray}
\left(1+g^2r^2_h\right)r^{D-3}_h+g^2qr^2_h-m=0.\label{ehrad:eq}
\end{eqnarray}
This equation has a unique solution corresponding to the event horizon, as a result of the presence of the non-trivial scalar field \cite{Cai2021}. The black hole temperature or the Hawking temperature is computed via the surface gravity $\kappa=f'(r_h)/(2\sqrt{H(r_h)})$ at the event horizon as follows
\begin{eqnarray}
T=\frac{\kappa}{2\pi}=\frac{f'(r_{h})}{4\pi\sqrt{H(r_h)}}.\label{eqn:temp}
\end{eqnarray}
The Bekenstein-Hawking entropy is determined by the area of the event horizon as follows
\begin{eqnarray}
S=\frac{\mathcal{A}}{4}=\frac{\omega_{D-2}}{4}\sqrt{H(r_h)}r_{h}^{D-2},\label{BHent:eq}
\end{eqnarray}
where $\omega_{D-2}=2\pi^{(D-1)/2}/\Gamma(\frac{D-1}{2})$ is the surface area of the unit $(D-2)$-sphere with $\Gamma(z)$ to be the Gamma function. 

The ADM mass $M$ and the charge $Q$ of black holes are calculated by using the conserved charges, leading to \cite{Lu2012}
\begin{eqnarray}
    M&=&\frac{\omega_{D-2}}{16\pi}\left[(D-2)m+(D-3)q\right],\label{ADM:mass}\\
    Q&=&\frac{(D-3)\omega_{D-2}}{16\pi}\sqrt{q(m+q)}.\label{eqn:charge}
\end{eqnarray}
The chemical potential $\Phi$ which is the conjugate quantity of the black hole charge reads
\begin{eqnarray}
    \Phi&=&\phi(\infty)-\phi(r_h)\nonumber\\
     &=&\left[1-\frac{1}{H(r_h)}\right]\sqrt{1+\frac{m}{q}}.\label{eqn:potential}
\end{eqnarray}
Additionally, we observe from Eqs. (\ref{Hr:exp}) and (\ref{fr:exp}) that the KK coupling constant $g$ behaves as a negative cosmological constant corresponding to the AdS curvature radius $L=1/g$. In this way, we define the cosmological constant or the KK coupling constant as the thermodynamic pressure in the extended phase space as follows
\begin{eqnarray}
    P&=&\frac{(D-1)(D-2)}{16\pi L^2}\nonumber\\
    &=&\frac{(D-1)(D-2)}{16\pi}g^2.
\end{eqnarray}

We can check that the thermodynamic quantities satisfy the first law of the black hole thermodynamic in the extended phase space as follows
\begin{eqnarray}
    dM=TdS+\Phi dQ+\mathcal{V}dP,\label{firstlaw}
\end{eqnarray}
where $\mathcal{V}$ is the thermodynamic volume which is the conjugate variable of the pressure $P$ and is determined by
\begin{eqnarray}
 \mathcal{V}&=&\left(\frac{\partial M}{\partial P}\right)_{S,Q}\nonumber\\
 &=&\frac{\omega_{D-2}}{D-3}\left[1+\frac{D-1}{2(D-2)}\frac{q}{r^{D-3}_h}\right]r^{D-1}_h.
\end{eqnarray}
In addition, we can check that the thermodynamic quantities satisfy the following Smarr relation
\begin{eqnarray}
    M=\frac{D-2}{D-3}TS+\Phi Q+\frac{2}{D-3}\mathcal{V}P.
\end{eqnarray}

In the following, we shall investigate the phase structure of the black hole, the stability of the thermodynamic phases, and the corresponding phase transitions, based on the free energy landscape in both the canonical ensemble (with the black hole charge kept fixed) and the grand canonical ensemble (with the chemical potential kept fixed).

\section{The thermodynamic phases in the canonical ensemble}
\label{sec:CE}
In this section, we study the thermodynamic phases and phase transitions of black holes in the canonical ensemble with the black hole charge $Q$ kept fixed. The generalized free energy in this ensemble is given by
\begin{eqnarray}
F=M-\tau S=F(r_h,\tau),    
\end{eqnarray}
 where $\tau$ is the ensemble temperature. The generalized free energy $F$ is an extension of the on-shell free energy with the black hole temperature replaced by the ensemble temperature. The generalized free energy $F$ is a function of $r_h$ and $\tau$, which is consistent with the fact that the ensemble comprises all black hole states with the event horizon radius taking the value from zero to infinity. In this way, the event horizon radius is realized as an order parameter that characterizes the thermodynamic phases of black holes. Hence, the value of the event horizon radius can help us to distinguish the different thermodynamic phases of black holes. The canonical ensemble includes both on-shell black hole states (which satisfy the equations of motion) and off-shell (fluctuating) black hole states (which are not the solutions of the equations of motion) with the charge kept fixed.

In order to study the free energy landscape of black holes in the canonical ensemble, let us express $m$ and $q$ in terms of the event horizon radius $r_h$, the black hole charge $Q$, and the KK coupling constant $g$. From Eqs. (\ref{ehrad:eq}) and (\ref{eqn:charge}), we find
\begin{eqnarray}
q&=&\frac{r^{D-3}_h}{2}\left(\sqrt{1+\frac{4\hat{Q}^2r^{2(3-D)}_h}{1+g^2r^2_h}}-1\right),\label{qcharge}\\
m&=&\left(1+g^2r^2_h\right)r_{h}^{D-3}+g^2r_{h}^{2}q,\label{m-mass}
\end{eqnarray}
where $\hat{Q}\equiv16\pi Q/[(D-3)\omega_{D-2}]$. By substituting these expressions of $q$ and $m$ into
Eq. (\ref{ADM:mass}), the ADM mass of black holes becomes
\begin{eqnarray}
M=\frac{\omega_{D-2}}{32\pi}r^{D-3}_h\left\{D-1+(D-2)g^2r^2_h+\left[D-3+(D-2)g^2r^2_h\right]\sqrt{1+\frac{4\hat{Q}^2r^{2(3-D)}_h}{1+g^2r^2_h}}\right\}.\label{ADM-mass-2}
\end{eqnarray}
Analogously, the Bekenstein-Hawking entropy given in Eq. (\ref{BHent:eq}) becomes
\begin{eqnarray}
S=\frac{\omega_{D-2}}{4\sqrt{2}}\left[1+\sqrt{1+\frac{4\hat{Q}^2r^{2(3-D)}_h}{1+g^2r^2_h}}\right]^{\frac{1}{2}}r_{h}^{D-2}.\label{BH-ent-2}
\end{eqnarray}
From Eqs. (\ref{ADM-mass-2}) and (\ref{BH-ent-2}), we can determine the generalized free energy $F$ in terms of the event horizon radius $r_h$, the thermodynamic pressure $P$, and the black hole charge $Q$. 

We can find the value of the free energy of the thermal AdS space which is the pure radiation without the black hole by taking a limit of the zero event horizon radius, given by
\begin{eqnarray}
\lim_{r_h\rightarrow0}F=Q\neq0.
\end{eqnarray}
This value is essential to determine the Hawking-Page (HP) phase transition between the thermal AdS space and the black hole \cite{Page1983}.

The extremal points of the generalized free energy $F$ are determined by
\begin{eqnarray}
    \frac{\partial F}{\partial r_h}=0\ \ \rightarrow \ \ \tau=\left(\frac{\partial M}{\partial S}\right)_{Q,P}.
\end{eqnarray}
From the first law given in Eq. (\ref{firstlaw}), we see that the ensemble temperature at the extremal points of $F$ is equal to the Hawking temperature $T$ which is rewritten by replacing the expressions of $q$ and $m$ given in Eqs. (\ref{qcharge}) and (\ref{m-mass}) into Eq. (\ref{eqn:temp}) as follows
\begin{eqnarray}
T=\frac{1}{2\sqrt{2}\pi r_h}\left[D-3+g^2r^2_h\left(D-2+\sqrt{1+\frac{4\hat{Q}^2r^{2(3-D)}_h}{1+g^2r^2_h}}\right)\right]\left[1+\sqrt{1+\frac{4\hat{Q}^2r^{2(3-D)}_h}{1+g^2r^2_h}}\right]^{-\frac{1}{2}}.
\end{eqnarray}
In this sense, the black hole solutions which satisfy the equations of motion given by Eqs. (\ref{Eins-eqs}), (\ref{vect-eqs}), and (\ref{scal-eqs}) would correspond to the extremal points of the generalized free energy.

\begin{figure}[ht]
  \centering
  \includegraphics[width=0.4\linewidth]{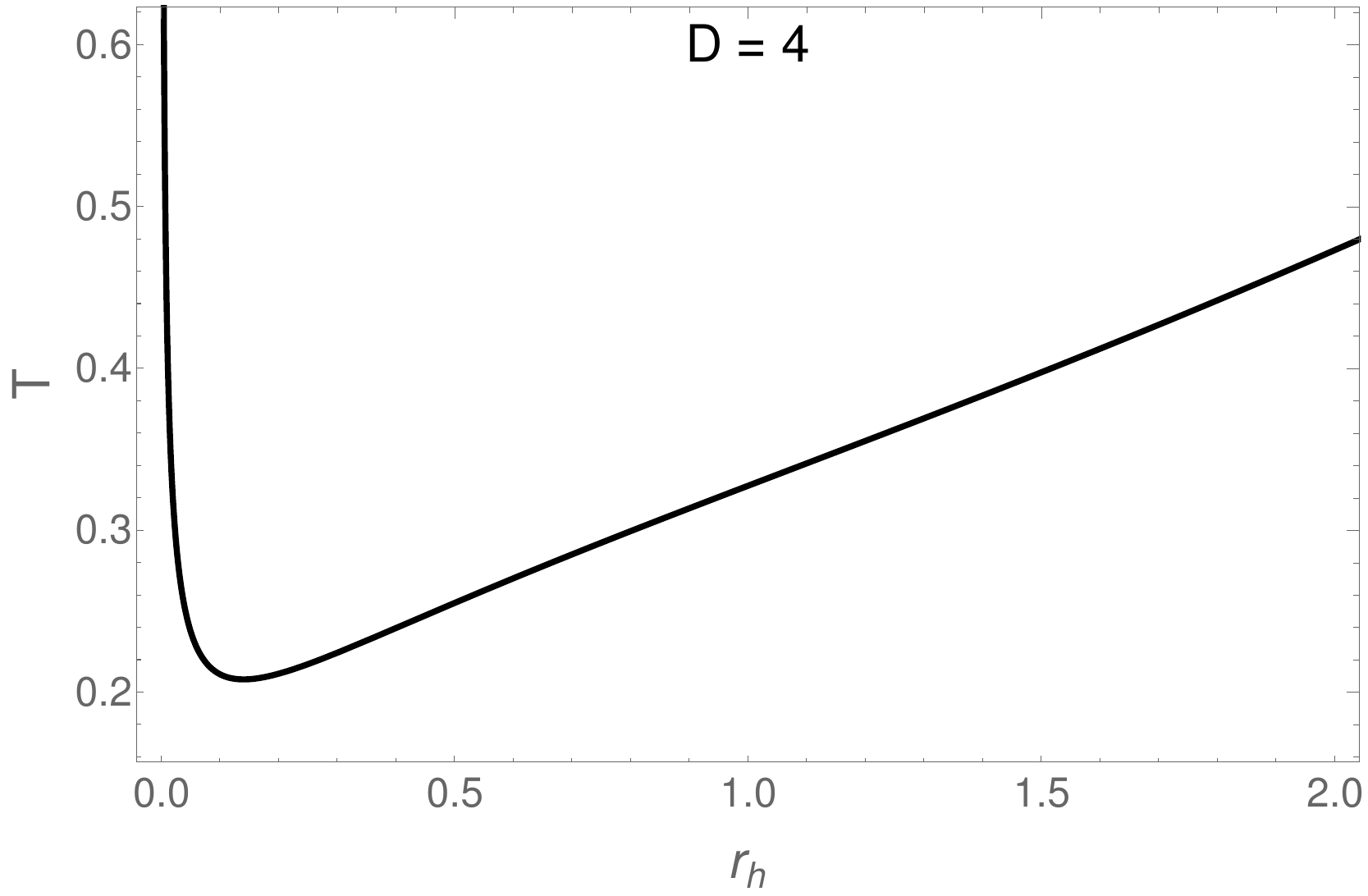}
  \includegraphics[width=0.4\linewidth]{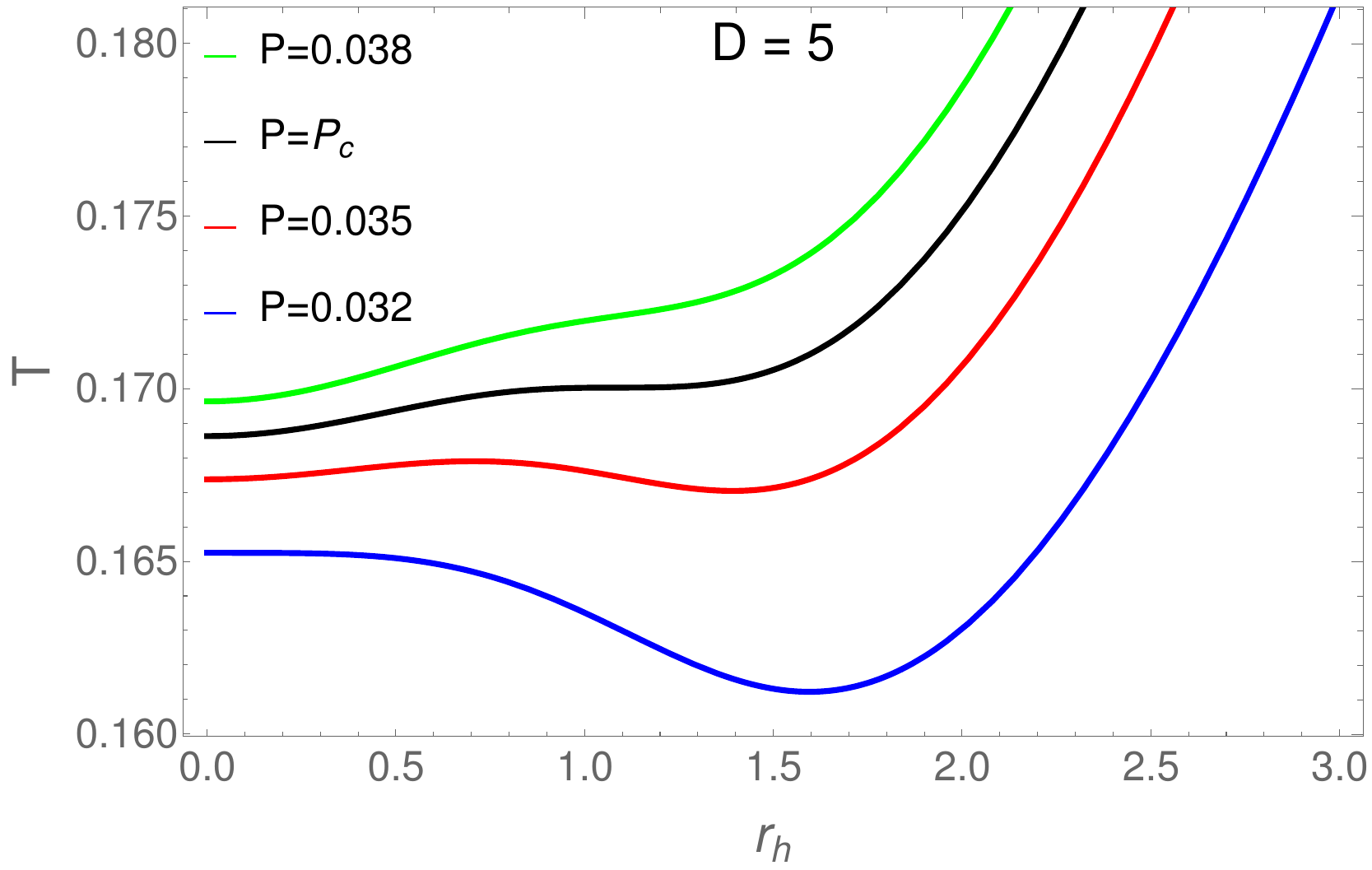}
  \includegraphics[width=0.4\linewidth]{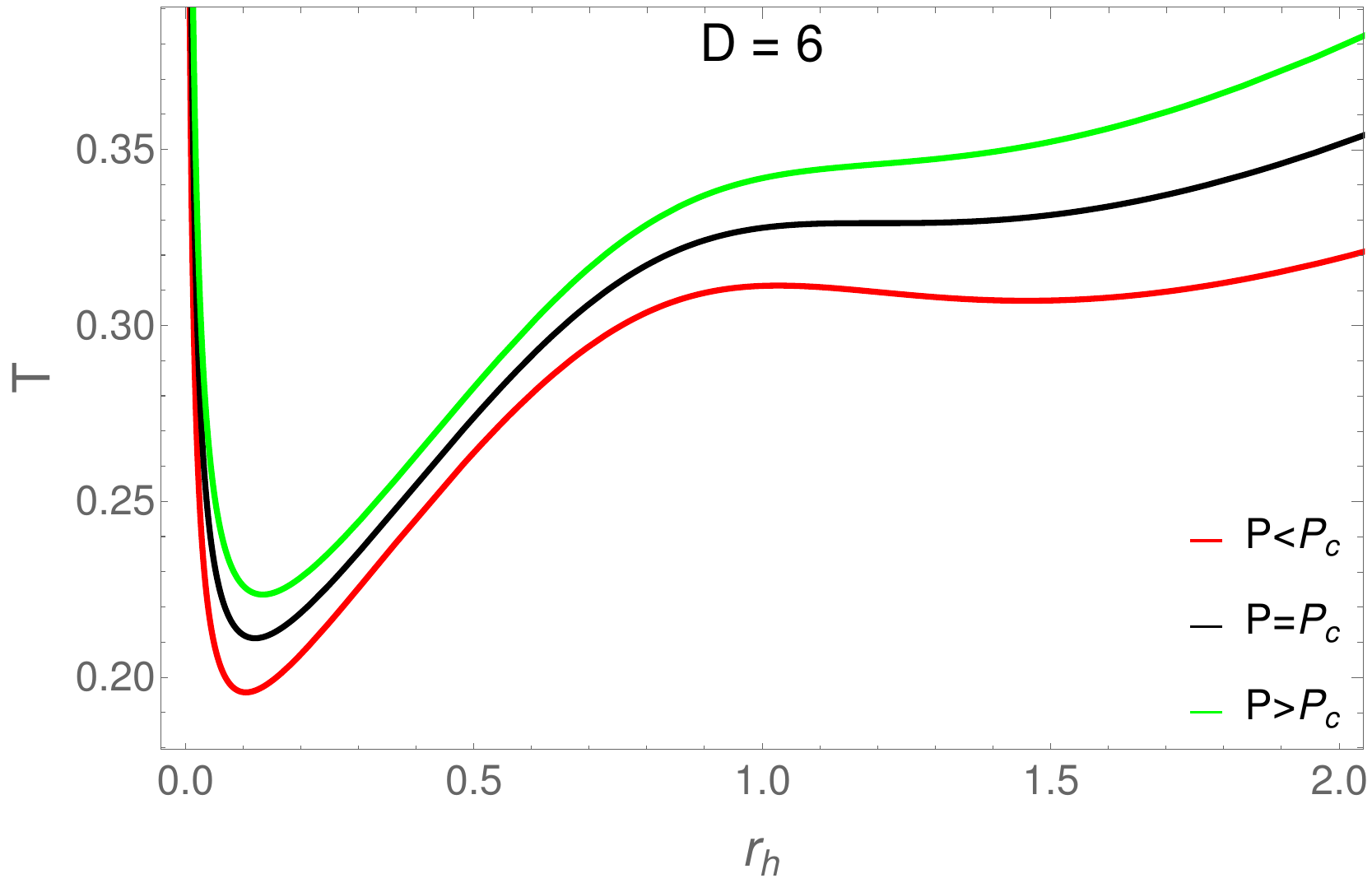}
  \includegraphics[width=0.4\linewidth]{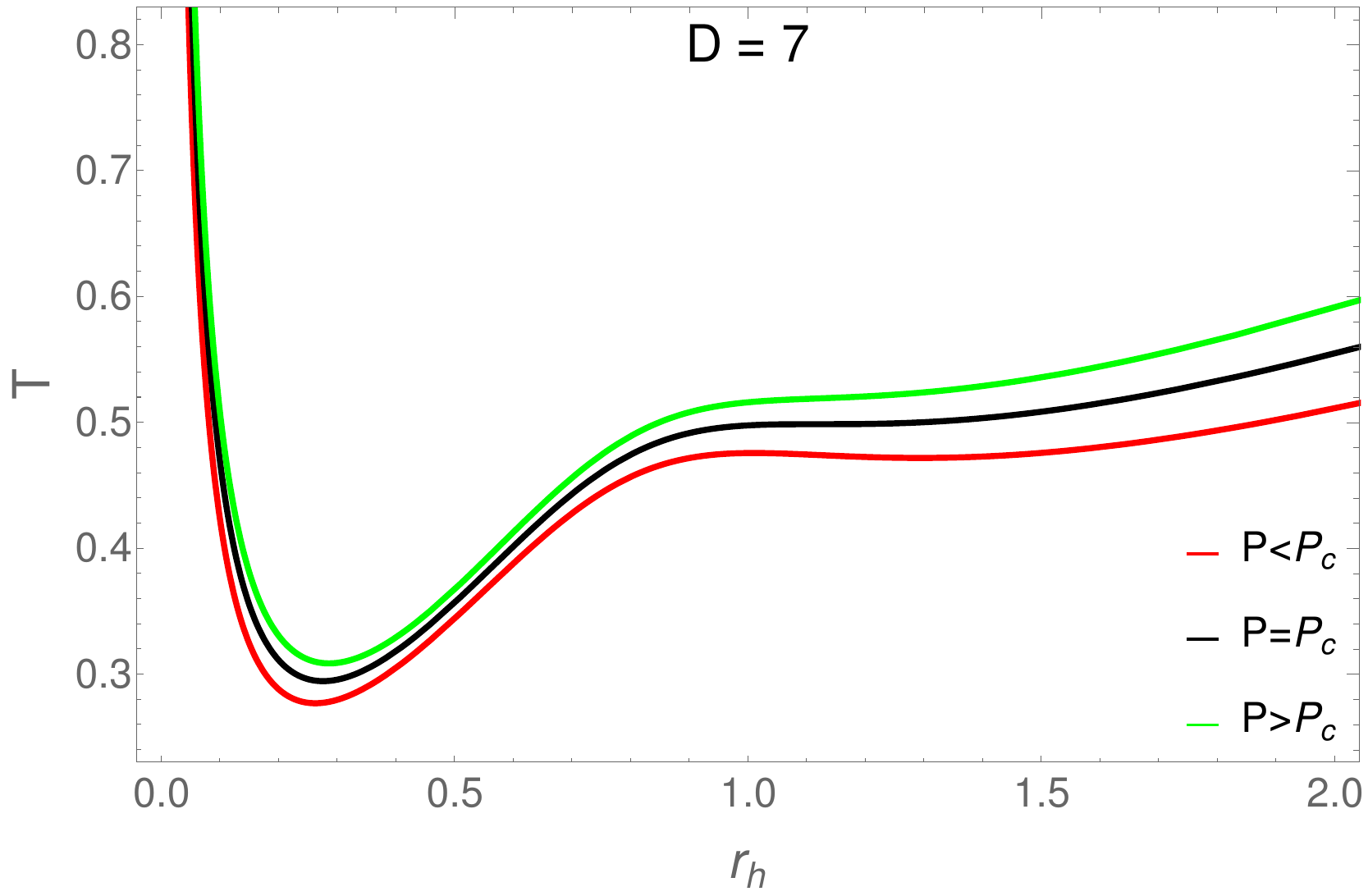}
  \caption{The behavior of the Hawking temperature in terms of the event horizon radius for the spacetime dimension $D=4$, $5$, $6$, and $7$ with various values of the pressure.}
  \label{fig:temp}
\end{figure}

We depict the behavior of the Hawking temperature in terms of the event horizon radius for various values of the spacetime dimension and the pressure in Fig. \ref{fig:temp}, with $Q$ kept fixed to be a unit (in the following discussions) for simplicity.
The behavior of the Hawking temperature $T$ depends strongly on the spacetime dimension and the pressure, which can be classified into the following three classes. The first class corresponds to $D=4$ where the isobaric curves $T-r_h$ have only one extremum independent of the value of the pressure. The second class corresponds to $D=5$ where the Hawking temperature approaches a constant in the limit of $r_h\rightarrow0$, given by
\begin{eqnarray}
\lim_{r_h\rightarrow0}T=\frac{1+8\pi g Q/\omega_3}{4\pi\sqrt{2\pi Q/\omega_3}}.    
\end{eqnarray}
In this way, black holes with the event horizon radius which is enough small have nearly the same temperature with the pressure kept fixed. In addition, when the pressure is above a critical value $P_c$, the isobaric curves $T-r_h$ do not have any extremum. On the contrary, when the pressure is lower than $P_c$, the isobaric curves exhibit some extrema, which implies the different thermodynamic phases of black holes and the phase transitions between them as seen later. The third class corresponds to $D=6$ and $D=7$. The behavior of the isobaric curves in this class is analogous to that of the first class when the pressure is higher than a critical value. When the pressure is below the critical value, the isobaric curves exhibit the addition of two extrema in analogy to the second class.

The critical point of the isobaric curves is determined by 
\begin{eqnarray}
    \frac{\partial T}{\partial r_h}=0,\quad \frac{\partial^2 T}{\partial r_{h}^{2}}=0.
\end{eqnarray}
By solving numerically, we find the critical values for the event horizon radius, the pressure, and the temperature, given in Table 1.
\begin{table}[h]
\begin{tabular}{|c c c c|} 
 \hline
 $D$ \quad \quad & $r_c$ \quad \quad & $P_c$ \quad \quad & $T_c$ \quad \quad\\ 
 \hline
 5\quad \quad & 1.091\quad \quad & 0.037\quad \quad & 0.170\quad \quad \\ 
 \hline
 6\quad \quad & 1.199\quad \quad & 0.116\quad \quad & 0.329\quad \quad \\
 \hline
 7\quad \quad & 1.120\quad \quad & 0.247\quad \quad & 0.498\quad \quad \\[0.5ex]
 \hline
\end{tabular}
\caption{Table 1. The numerical values for the critical event horizon radius $r_c$, the critical pressure $P_c$, and the critical temperature $T_c$ with $5\leq D\leq 7$.}
\end{table}
The $P-V$ criticality of black holes has been studied in \cite{HaoWei2019}.  

\subsection{The first class: $D=4$}
\begin{figure}[ht]
% Requires \usepackage{graphicx}\
 \centering
\begin{tabular}{cc}
\includegraphics[width=0.5 \textwidth]{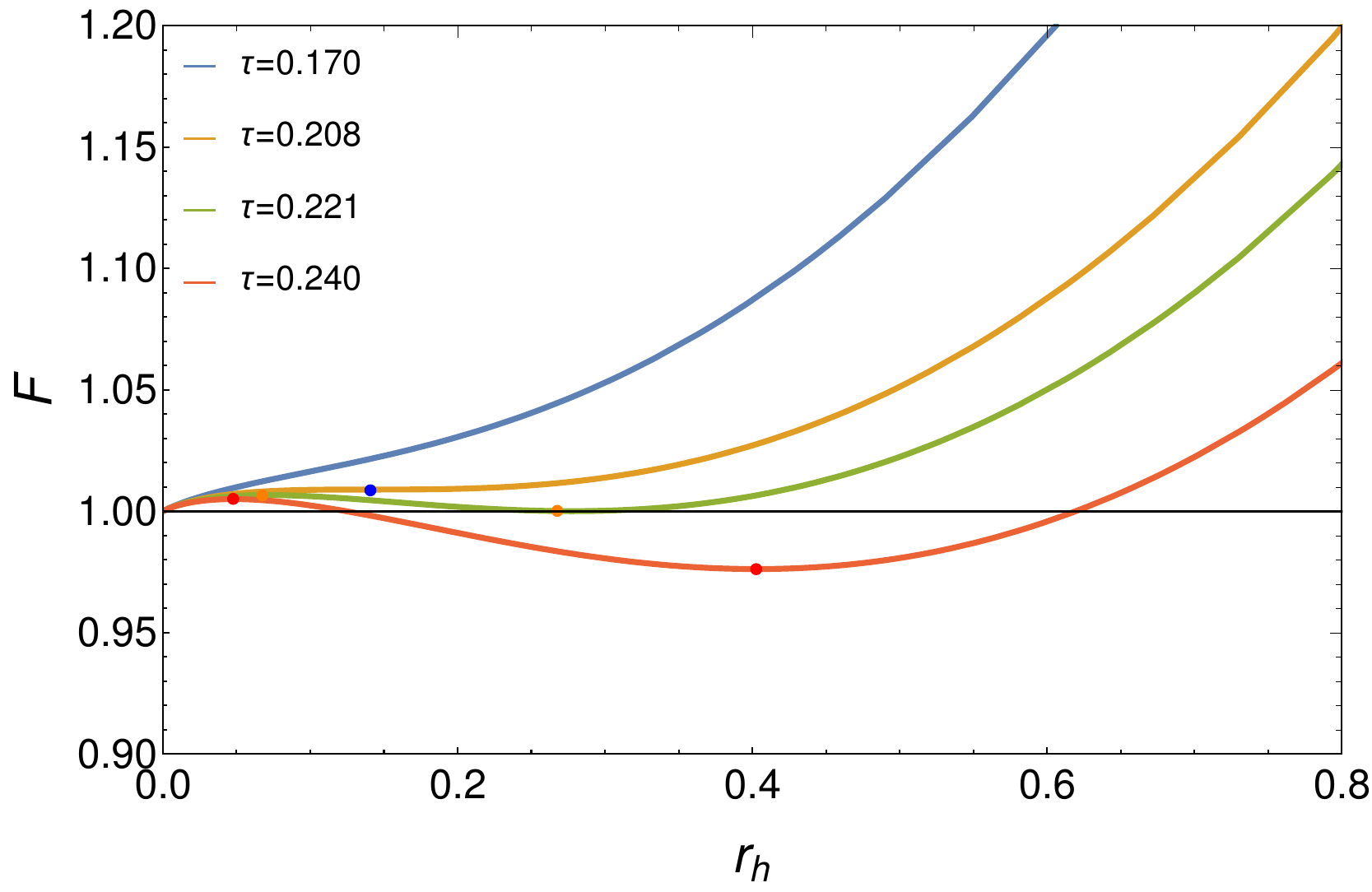}
\end{tabular}
 \caption{The generalized free energy $F$ is plotted in the event horizon radius $r_h$ with the change of the ensemble temperature $\tau$ for $D=4$ and $P=0.1$. The blue point refers to the inflection point. The orange and red points represent the extremum.}\label{genfreeen-D4-plot}
\end{figure}

\begin{figure}[ht]
% Requires \usepackage{graphicx}\
 \centering
\begin{tabular}{cc}
\includegraphics[width=0.5 \textwidth]{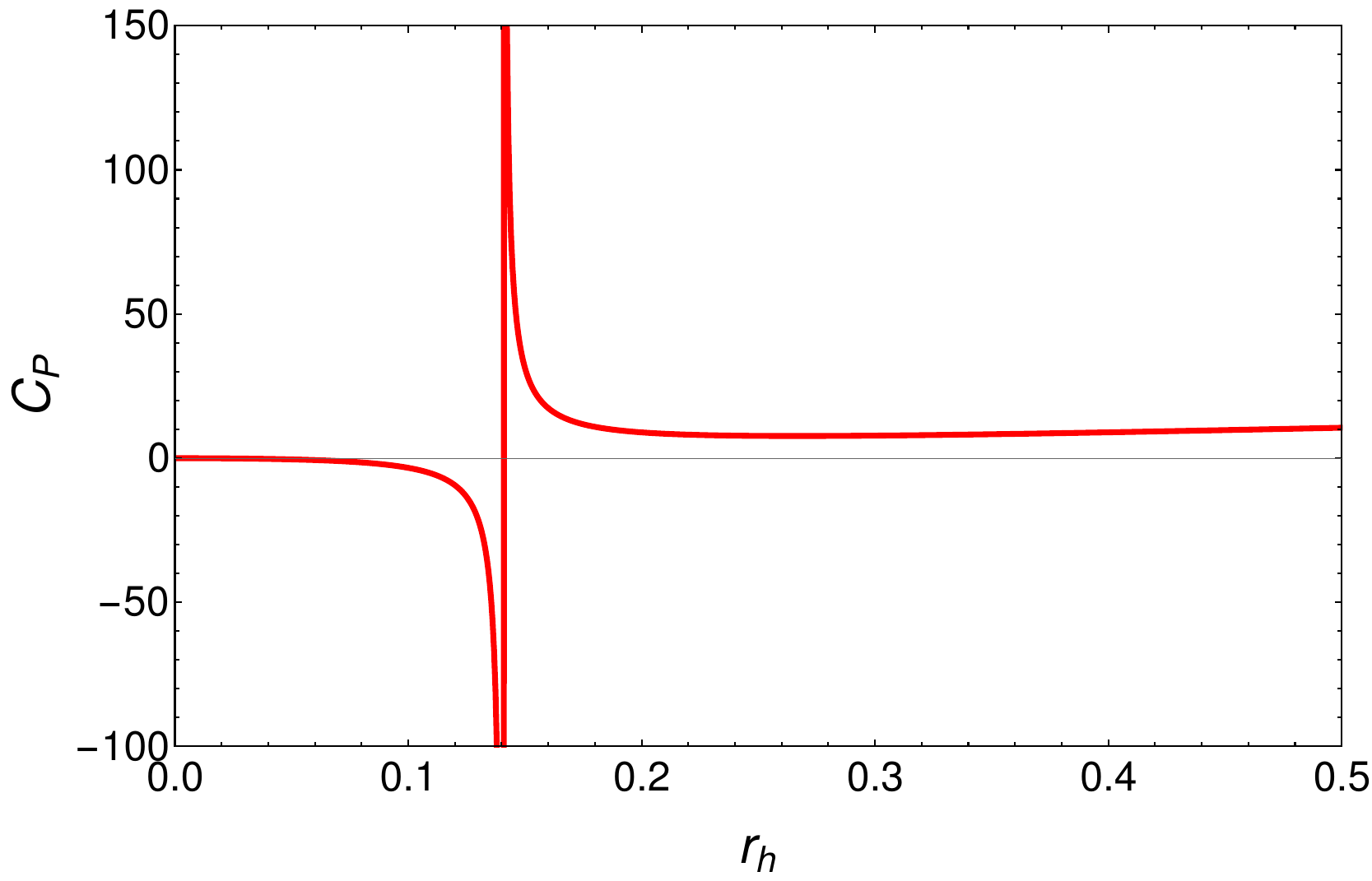}
\end{tabular}
 \caption{The heat capacity at the constant pressure in terms of the event horizon radius at $D=4$.}\label{heatcap-D4-plot}
\end{figure}

\begin{figure}[ht]
% Requires \usepackage{graphicx}\
 \centering
\begin{tabular}{cc}
\includegraphics[width=0.4 \textwidth]{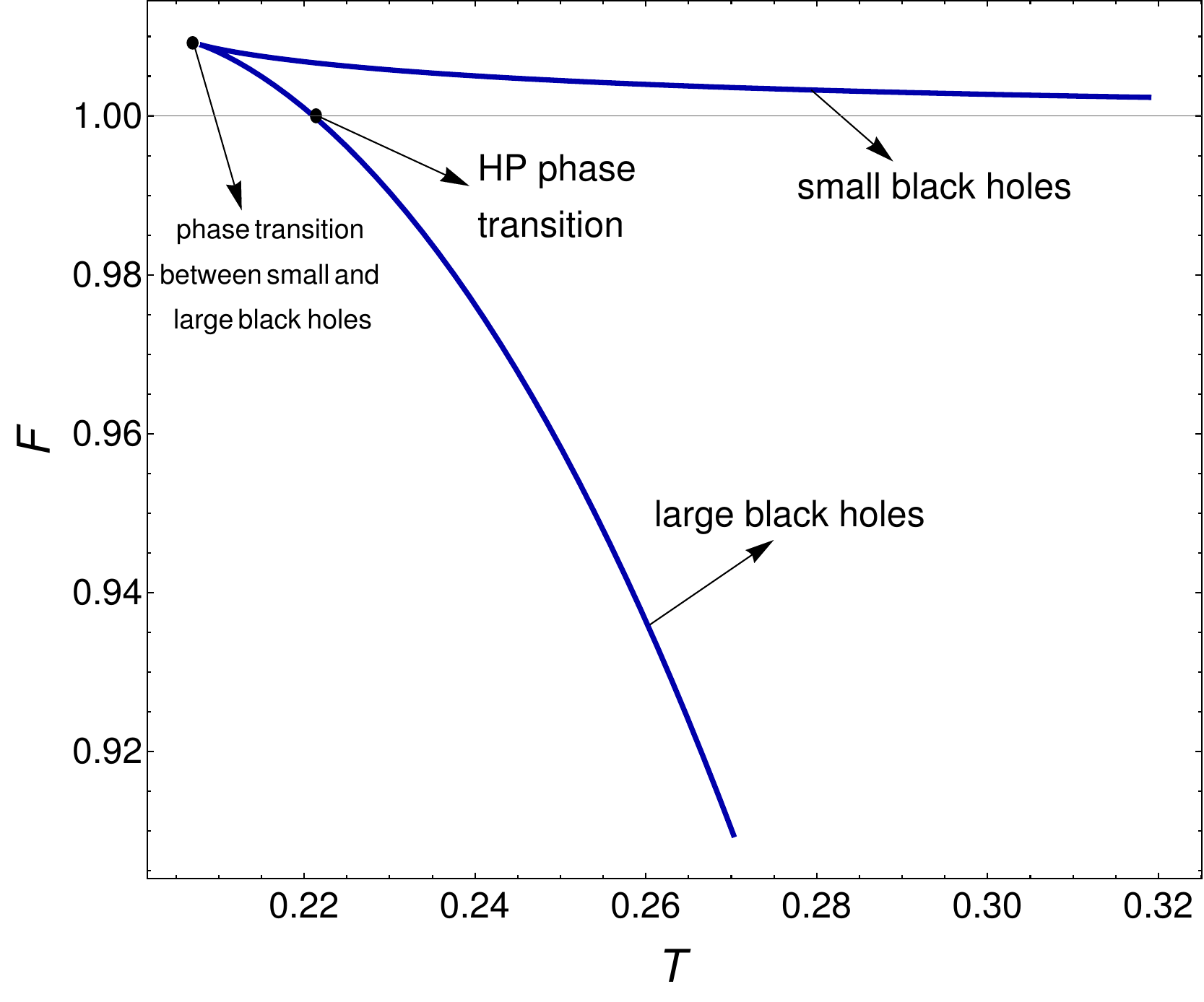}
\end{tabular}
 \caption{The on-shell free energy as a function of the black hole temperature for $D=4$ and $P=0.1$.}\label{onsh-freenr-D4-plot}
\end{figure}

We study the landscape of the generalized free energy $F$ as the function of the event horizon radius for various values of the ensemble temperature $\tau$ in the case of $D=4$ in Fig. \ref{genfreeen-D4-plot}.
When the ensemble temperature is lower than a minimal value which is $\approx0.208$ for $P=0.1$, the generalized free energy $F$ has only one global minimum with $F=1$ located at the origin $r_h=0$. This implies that the system corresponds to the thermal AdS space. When the ensemble temperature is above the minimal value, the generalized free energy $F$ exhibits one minimum and a local maximum corresponding to the thermodynamically unstable and stable black holes, respectively. We can also point to the thermodynamic stability of these black hole phases based on investigating the behavior of the heat capacity at the constant pressure defined by
\begin{eqnarray}
C_P &\equiv& T\left(\frac{\partial S}{\partial T}\right)_P=T\left(\frac{\partial S}{\partial r_h}\right)_P\left(\frac{\partial T}{\partial r_h}\right)_P^{-1}.
\end{eqnarray}
As depicted in Fig. \ref{heatcap-D4-plot}
we observe that the large black holes corresponding to the minimum of the generalized free energy have a positive heat capacity, hence they are thermodynamically stable. Whereas, the small black holes the local maximum of the generalized free energy have a negative heat capacity, implying that they are thermodynamically unstable. This means there is a second-order phase transition from the small black holes to the large black holes, corresponding to the divergence of $C_P$. We also indicate this phase transition from the multivalued behavior of the on-sell free energy, as depicted in Fig. \ref{onsh-freenr-D4-plot}.
The upper branch represents the small black holes with the free energy that is higher than that of the large black holes represented by the lower branch. The intersection of these two branches determines the phase transition between the small and large black holes.

In addition, from Figs. \ref{genfreeen-D4-plot} and \ref{onsh-freenr-D4-plot}, we observe that when the temperature is above a value $\approx0.221$ (for $P=0.1$), the free energy of the large black holes is lower than that of the thermal AdS space. On the other hand, the large black holes correspond to the global minimum of the free energy. As a result, the large black holes are the dominant contribution to the partition function, hence they are the thermodynamically preferred states. This indicates the HP phase transition from the thermal radiation state to the large black holes. The temperature of the HP phase transition is easily determined by the intersection between the branch of the large black holes and the horizontal line $F=1$, in Fig. \ref{onsh-freenr-D4-plot}.

\subsection{The second class: $D=5$}
Unlike the case of $D=4$, the behavior of the generalized free energy $F$ for $D=5$ depends on not only the ensemble temperature $\tau$ but also the thermodynamic pressure $P$ generated by gauging the KK gauge field. The phase structure and the phase transitions in $D=5$ thus are richer than those in $D=4$. 

\begin{figure}[ht]
 \centering
  \includegraphics[width=0.4\linewidth]{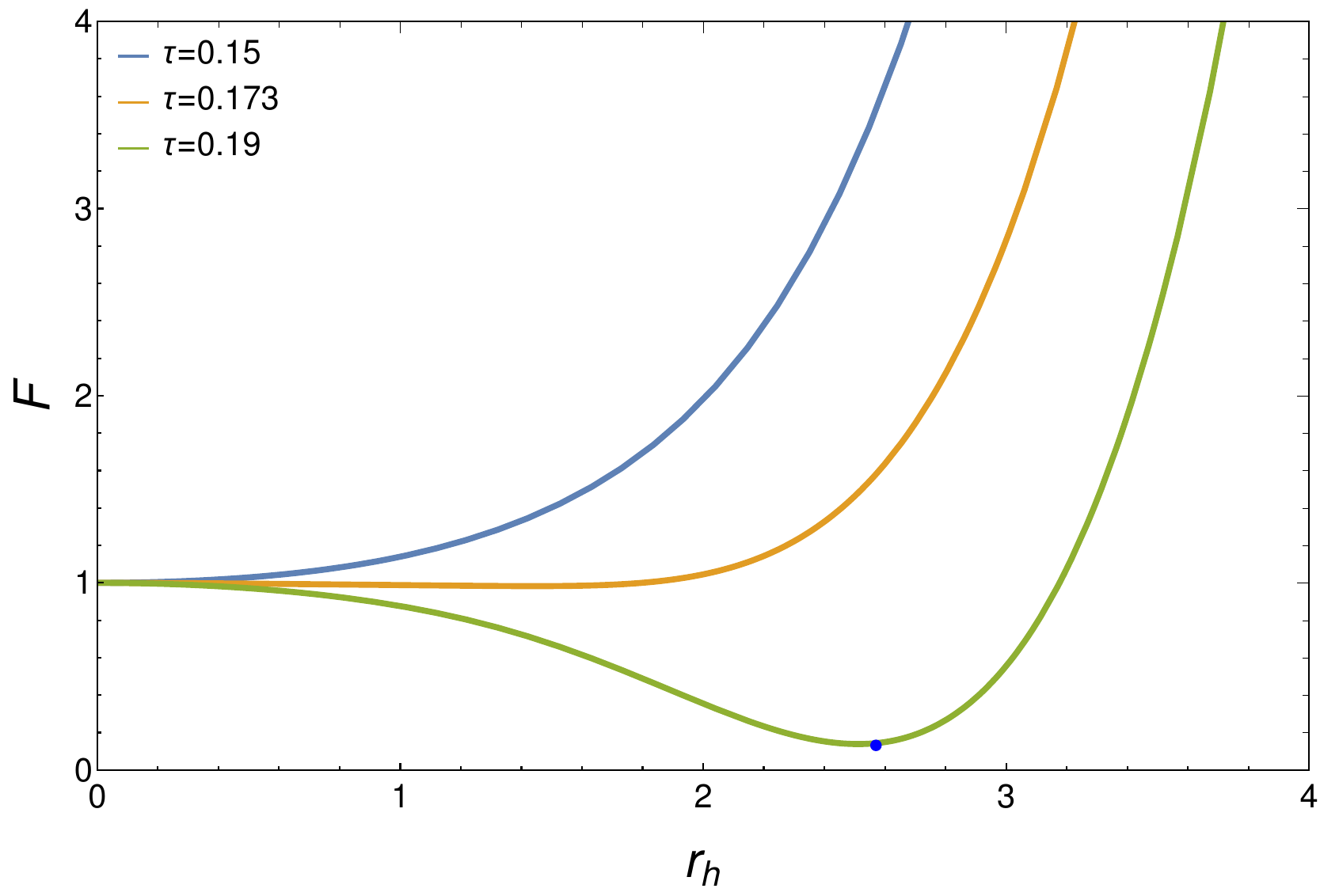}
  \caption{The generalized free energy $F$ versus the event horizon radius $r_h$ for $D=5$ and $P=0.038$.}
  \label{gefreeenr-D5-01-plot}
\end{figure}

\begin{figure}[ht]
  \centering
  \includegraphics[width=0.4\linewidth]{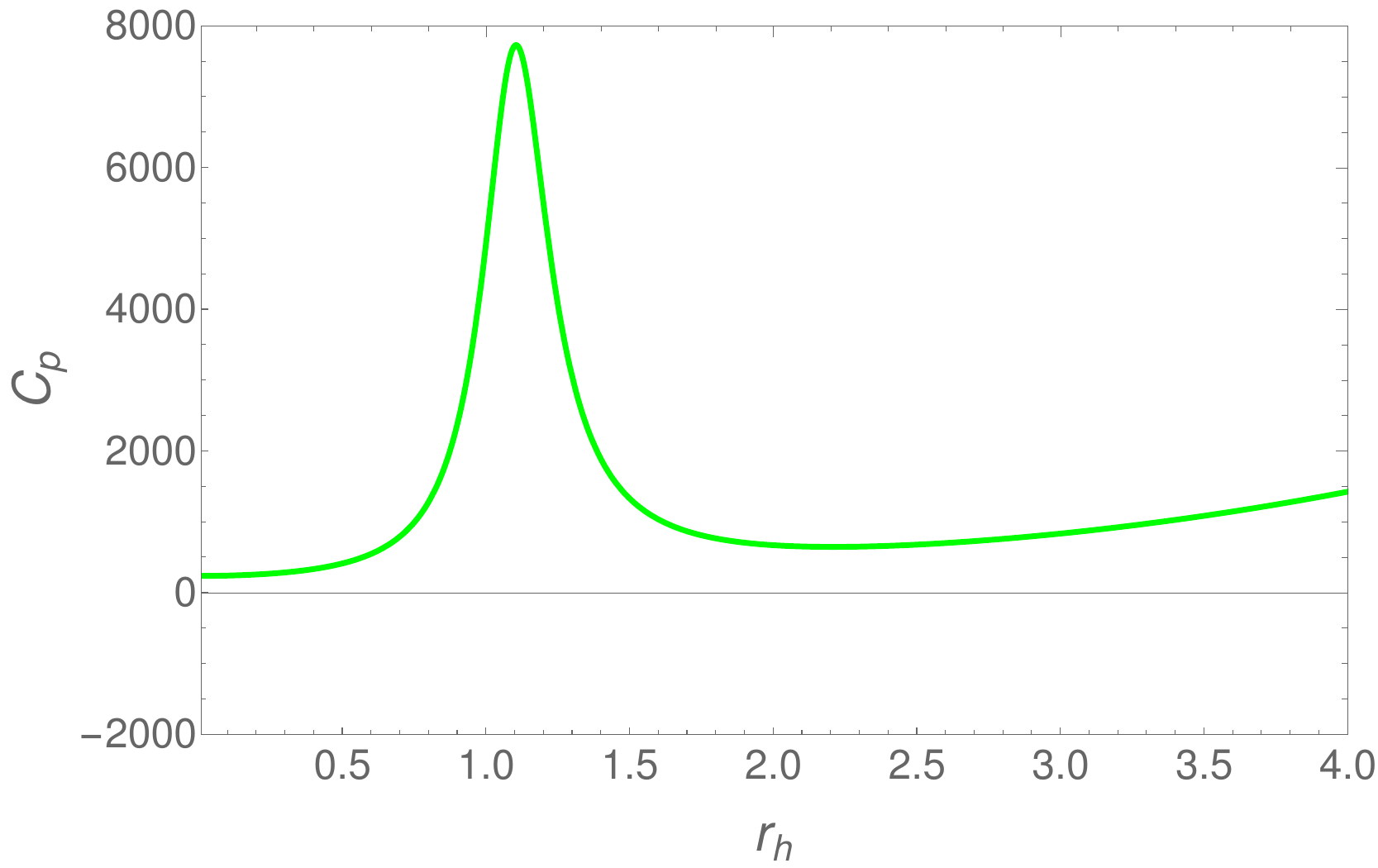}
  \includegraphics[width=0.4\linewidth]{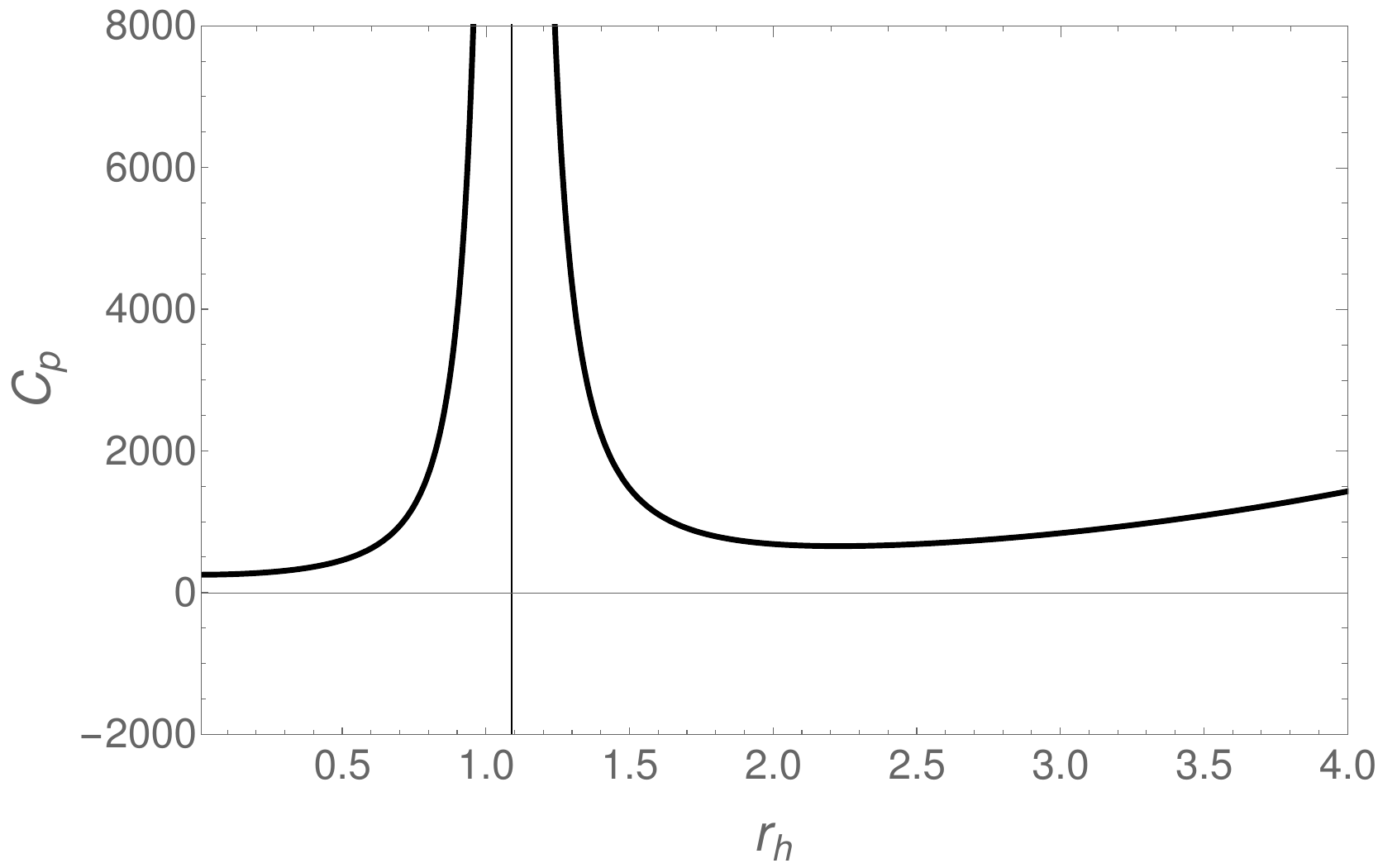}\\
  \includegraphics[width=0.4\linewidth]{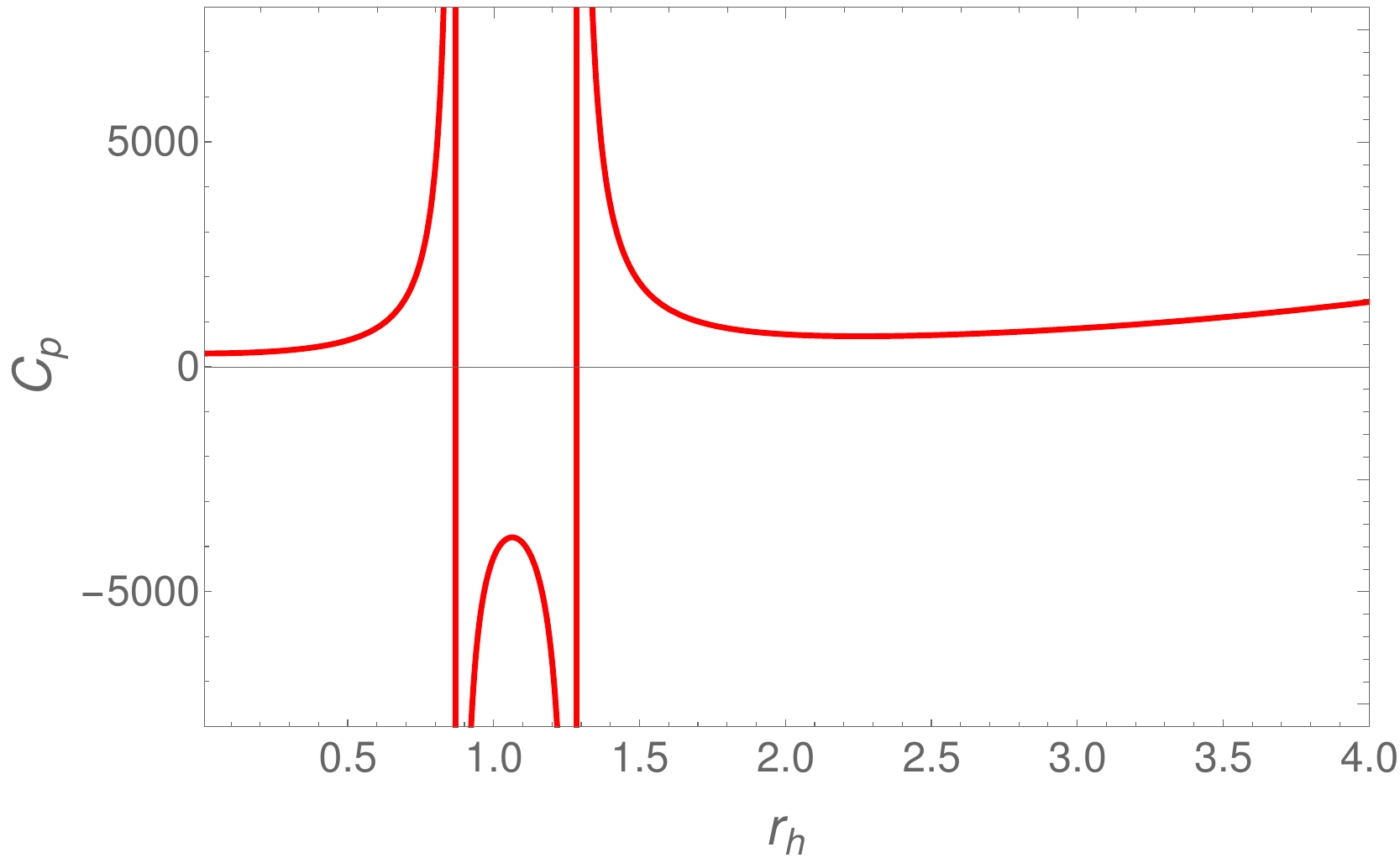}
  \caption{The behavior of the heat capacity at the constant pressure $C_P$ in terms of the event horizon radius for three cases: $P>P_c$ (top-left), $P=P_c$ (top-right), and $P<P_c$ (bottom).}
  \label{fig:cp5}
\end{figure}

For the pressure $P>P_c\approx0.037$, we depict the behavior of the generalized free energy $F$ in terms of the event horizon radius for various values of the ensemble temperature $\tau$ in Fig. \ref{gefreeenr-D5-01-plot}.
When $\tau$ is lower than a minimum value $\tau_{\text{m}}$ which is about $0.173$ for $P=0.038$, the generalized free energy possesses only a minimum with $F=1$ located at $r_h=0$, corresponding to the pure radiation state or the thermal AdS space. Increasing the ensemble temperature such that $\tau\gtrsim \tau_{\text{m}}$ leads to the occurrence of a global minimum (represented by the blue dot in Fig. \ref{gefreeenr-D5-01-plot}) in the curve of the generalized free energy, corresponding to the presence of the on-shell black hole states. Because the free energy of black holes is lower than that of the thermal AdS space and they have a positive heat capacity as seen in the top-left panel of Fig. \ref{fig:cp5}, black holes are the thermodynamically preferred states. 

\begin{figure}[ht]
 \centering
  \includegraphics[width=0.4\linewidth]{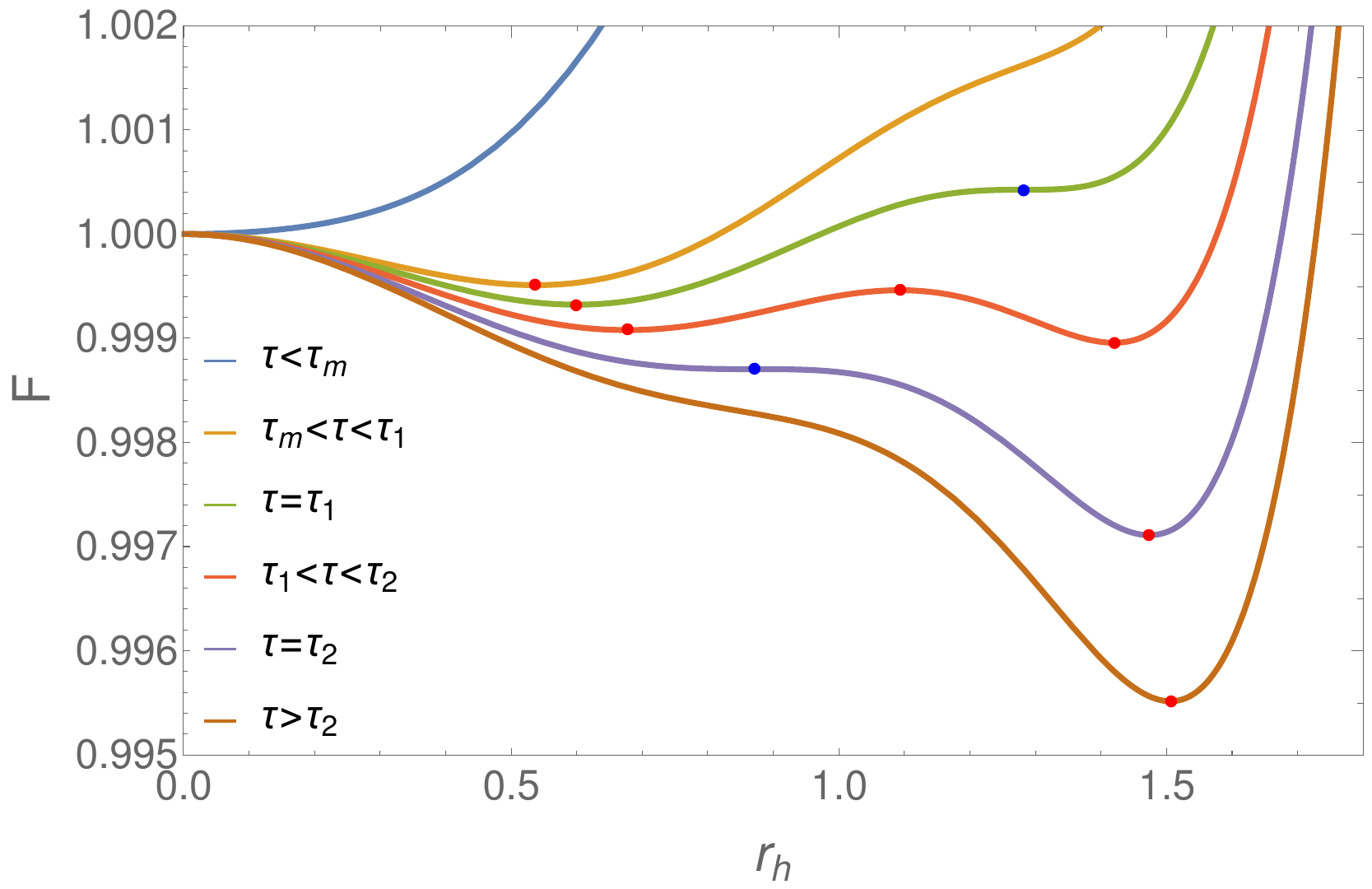}
  \includegraphics[width=0.4\linewidth]{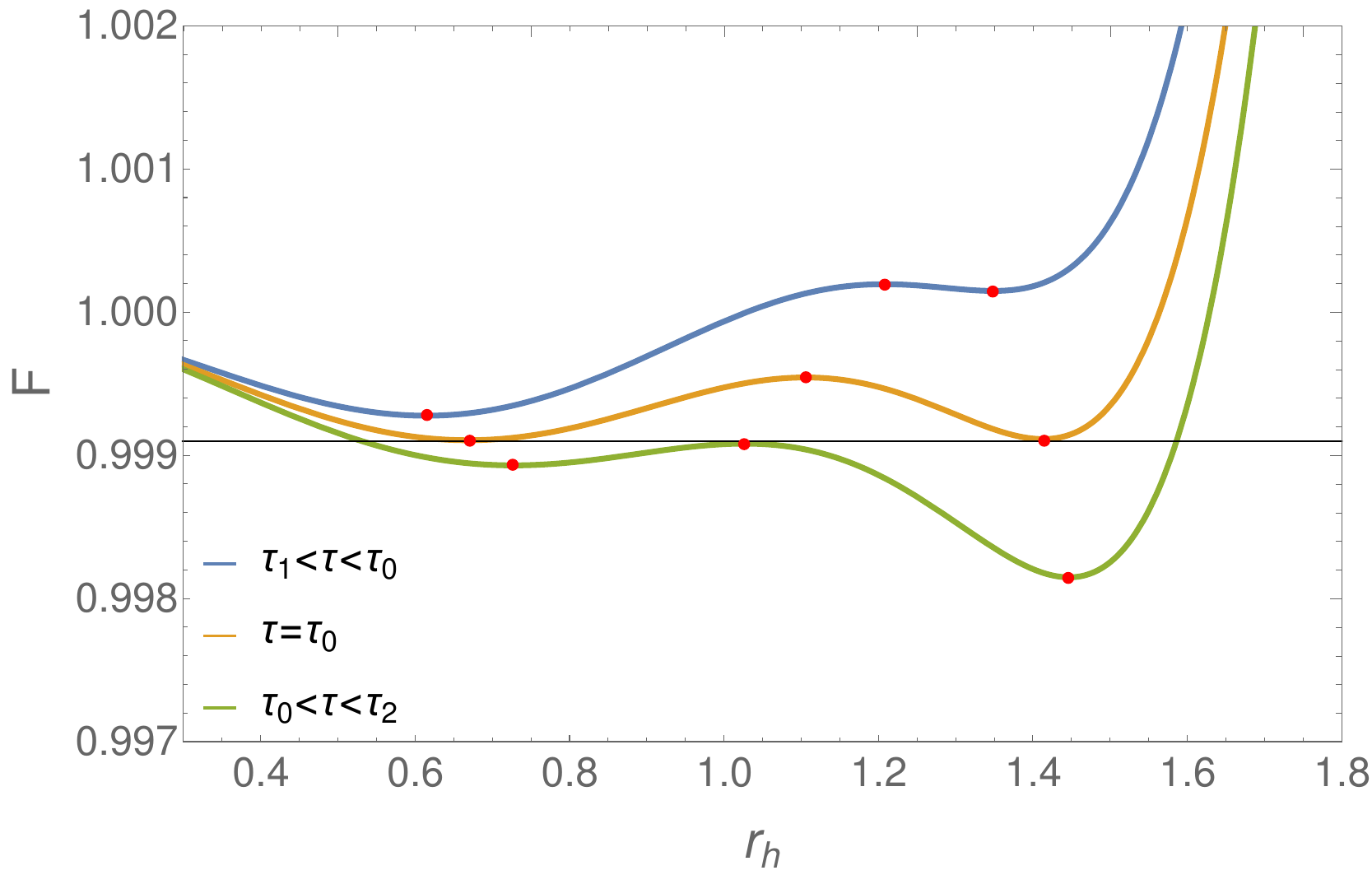}
  \caption{The behavior of the generalized free energy $F$ in the event horizon radius $r_h$ for $D=5$ and $P=0.036$. $F$ has the shapes of a double well when $\tau_1<\tau<\tau_2$ with $\tau_1 \approx 0.1689 $ and $\tau_2 \approx 0.1691$, and a single well in other cases.  The red points are the local extremal points that represent the on-shell states. The blue points represent the inflection point. Two local minima have the same value when $\tau=\tau_0 \approx 0.16899$.}
  \label{fig:FD5}
\end{figure}

For the pressure $P<P_c$, we plot the generalized free energy $F$ as a function of the event horizon radius for different values of the ensemble temperature $\tau$ in Fig. \ref{fig:FD5}. 
In analogy to the case of $P>P_c$, the thermal AdS space exits when the ensemble temperature is lower than $\tau_{\text{m}}$ which is about $0.1683$ for $P=0.036$. However, some extremal points of the generalized free energy $F$ appear for $\tau>\tau_{\text{m}}$ which corresponds to the on-shell black hole states. 

We observe that there are two critical values of the ensemble temperature which are $\tau_1 \approx 0.1689 $ and $\tau_2 \approx 0.1691$ for $P=0.036$. When $\tau_1<\tau<\tau_2$, the generalized free energy $F$ possesses two local minima and one local maximum corresponding to the on-shell black hole states. The two local minima represent the small and large black holes which are thermodynamically stable because their heat capacity is positive as seen in the bottom panel of Fig. \ref{fig:cp5}.
Which one prefers thermodynamically depends on its relative depth in the free energy landscape. We can find a value of the ensemble temperature $\tau_0$ where the free energy of the small and large black holes has the same value. For $\tau<\tau_0$, the free energy of the small black holes is lower than that of the large black holes, indicating that the small black holes are the thermodynamically preferred states. On the contrary, for $\tau>\tau_0$, the large black holes are the global stable state and hence they are thermodynamically preferred. When the ensemble temperature is above $\tau_2$, the generalized free energy only manifests an extremum which is a global minimum. This means that there is a unique thermodynamic phase and hence there is no phase transition.

The local maximum of the generalized free energy $F$ represents the intermediate black holes that have a negative heat capacity (as seen in the bottom panel of Fig. \ref{fig:cp5}) which implies the thermodynamic instability. As a result, the intermediate black holes would decay into the small or large black holes, which leads to the second-order phase transitions. We find these second-order phase transitions by observing the divergence of the heat capacity. In the bottom panel of Fig. \ref{fig:cp5}, we observe that there are two divergences in the behavior of the heat capacity $C_P$ which divide the domain of black holes into three subdomains: the small black holes, the intermediate black holes, and the large black holes. The divergence with the small $r_h$ indicates the phase transition between the small black holes and the intermediate black holes, whereas the one with the large $r_h$ exhibits the phase transition between the large black holes and the intermediate black holes. For $\tau<\tau_0$, because the small black holes are preferred thermodynamically, the rate at which the intermediate black holes decay into the small black holes would be greater than the rate at which the intermediate black holes transform into the large black holes. However, this happens oppositely when $\tau>\tau_0$. 

\begin{figure}[ht]
  \includegraphics[scale=0.3]{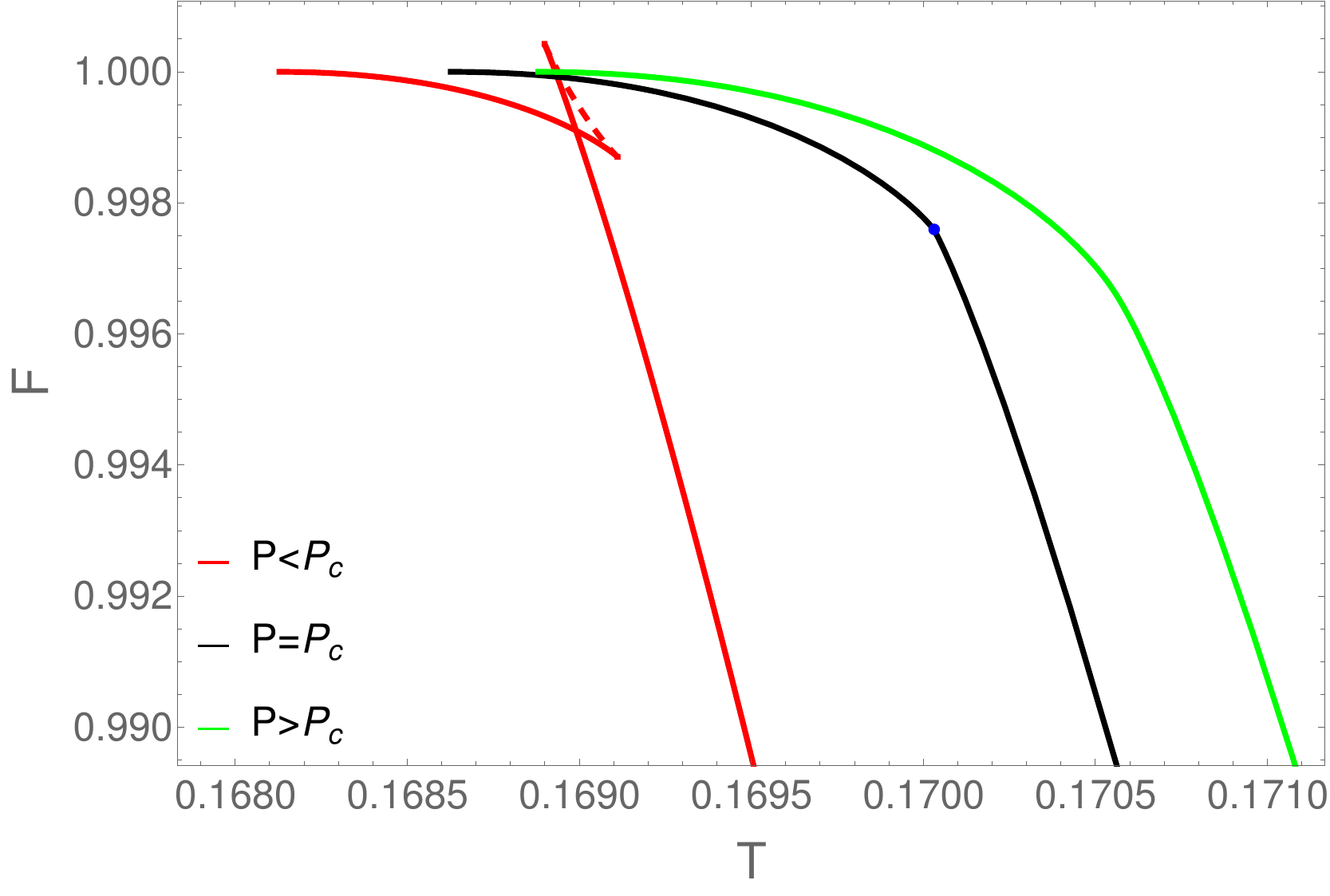}
  \caption{The on-shell free energy to the temperature for five-dimensional black holes in three cases: $P<P_{c}$ (red), $P=P_{c}$ (black), and $P>P_{c}$ (green). The blue dot is the critical point of the Van der Waals-like phase transition.}
  \label{fig:gibbs5}
\end{figure}

We can observe more phase transitions of black holes from the behavior of the on-shell free energy as depicted in Fig. \ref{fig:gibbs5}. 
The green, black, and red curves represent the on-shell free energy of black holes for $P>P_c$, $P=P_c$, and $P<P_c$, respectively. The dashed red line corresponds to the intermediate black holes, while the upper and lower branches of the solid red curve show the small and large black holes, respectively. The behavior of the on-shell free energy only reveals the phase transitions when $P<P_c$. We can observe the second-order phase transitions between the small(large) black holes and the intermediate black holes from the multivalued behavior of the on-sell free energy. Furthermore, the on-shell free energy exhibits the behavior of the \emph{swallowtail}, indicating the first-order phase transition between the small black holes and the large black holes because the difference of the entropy in the two phases is non-zero. This phase transition is analogous to the Van der Waals phase transition between the liquid and gas.

\subsection{The third class: $D=6$ and $7$}
The generalized free energy $F$ always has a local minimum and a local maximum which corresponds to the thermodynamically stable and unstable black holes, respectively, independent of the value of the pressure.
\begin{figure}[ht]
   \centering
  \includegraphics[width=0.45\linewidth]{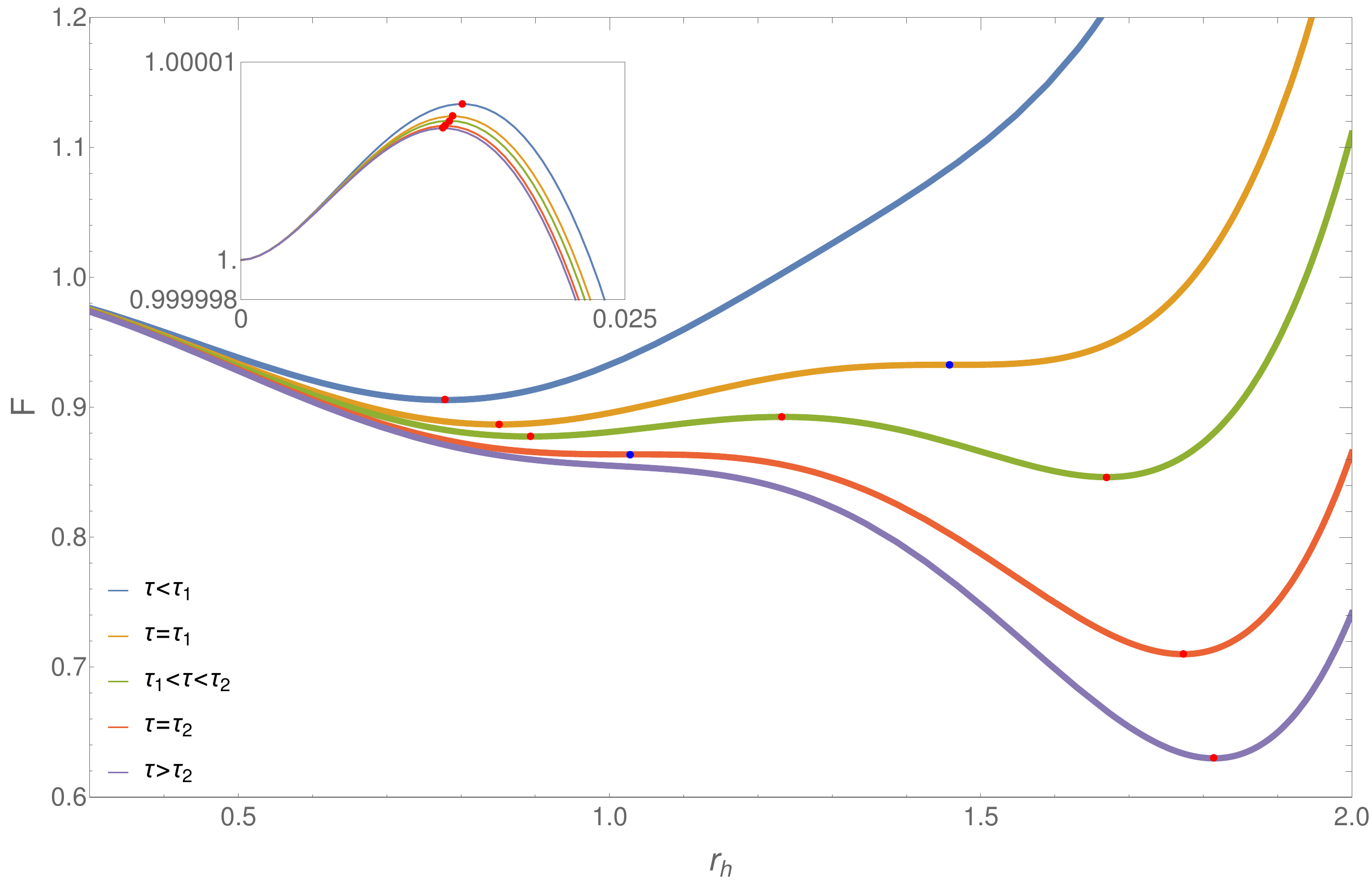}
  \includegraphics[width=0.45\linewidth]{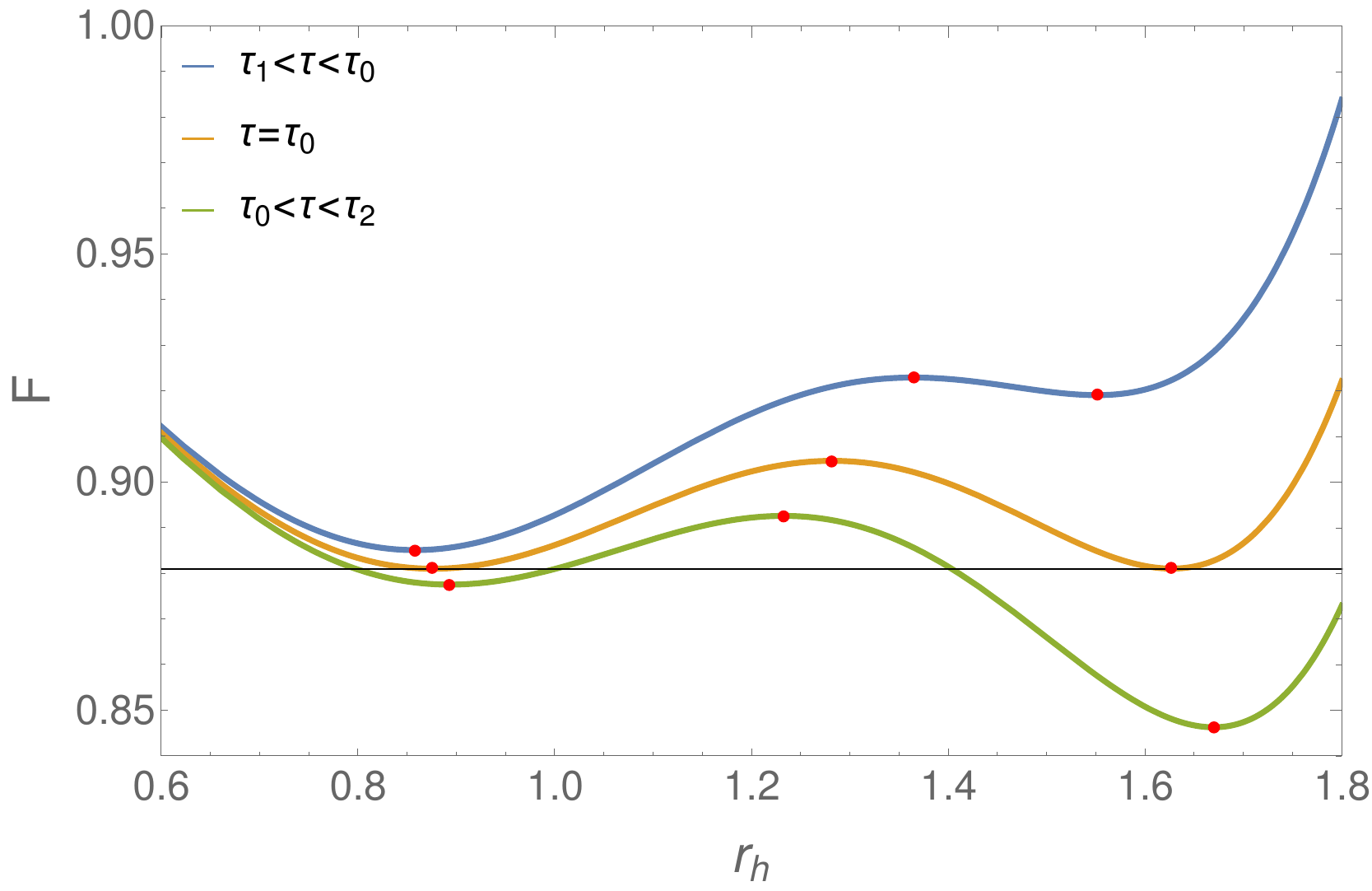}
  \caption{The behavior of the generalized free energy $F$ in the event horizon radius $r_h$ for $D=6$ with $P=0.1$. $F$ has a double well when $\tau_1<\tau<\tau_2$ with $\tau_1 \approx 0.3070$ and $ \tau_2 \approx 0.3113$, and a single well in other cases. The red points are the local extremal points that represent the on-shell black hole states. The blue points represent the inflection points. Two local minima have the same value when $\tau=\tau_0 \approx 0.3083$.}
  \label{fig:FD6}
\end{figure}
This particular point is shown by the small panel given in Fig. \ref{fig:FD6} and it can also be observed from the behavior of both heat capacity and on-shell free energy. 
\begin{figure}[ht]
  \centering
  \includegraphics[width=0.4\linewidth]{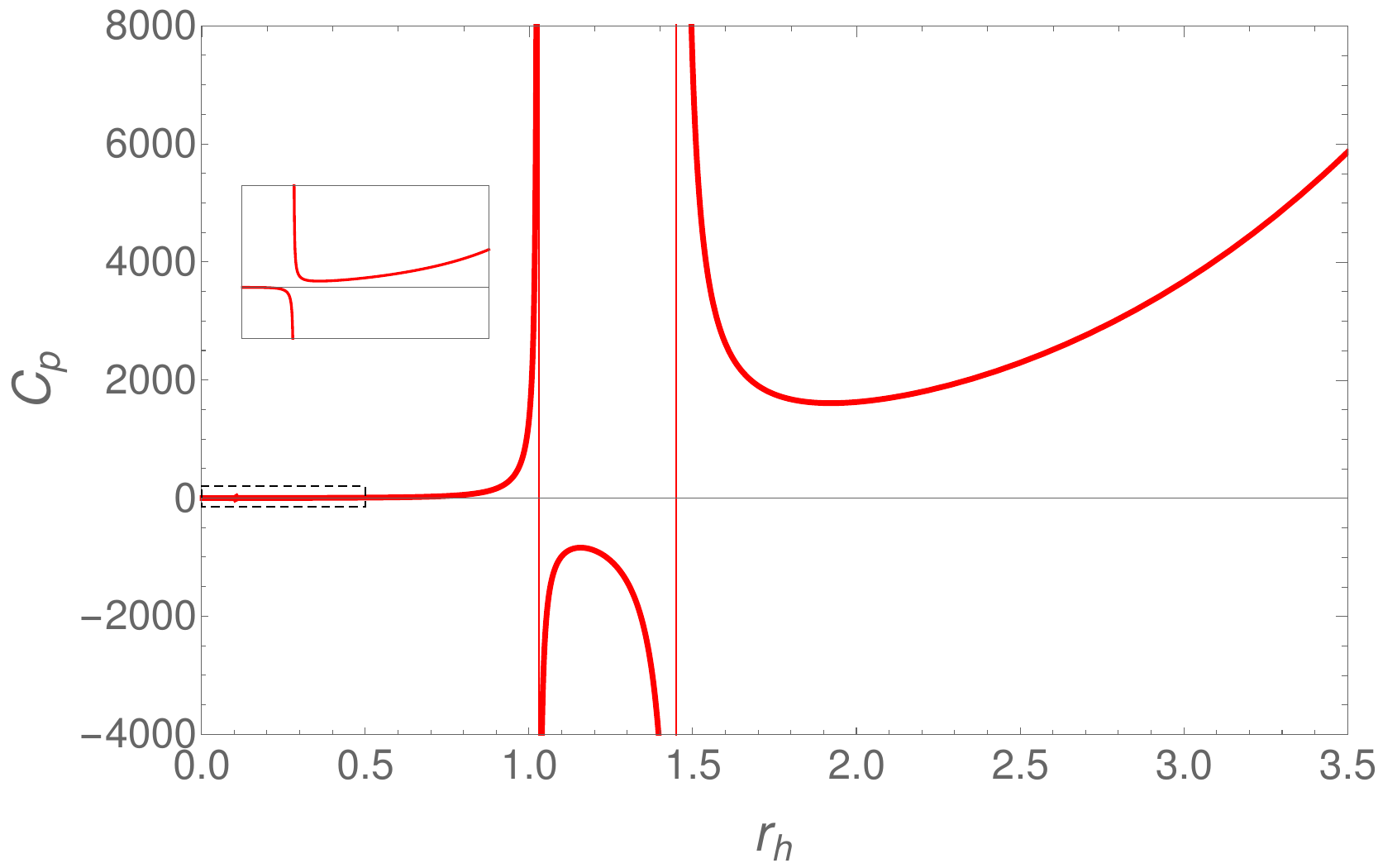}
  \includegraphics[width=0.4\linewidth]{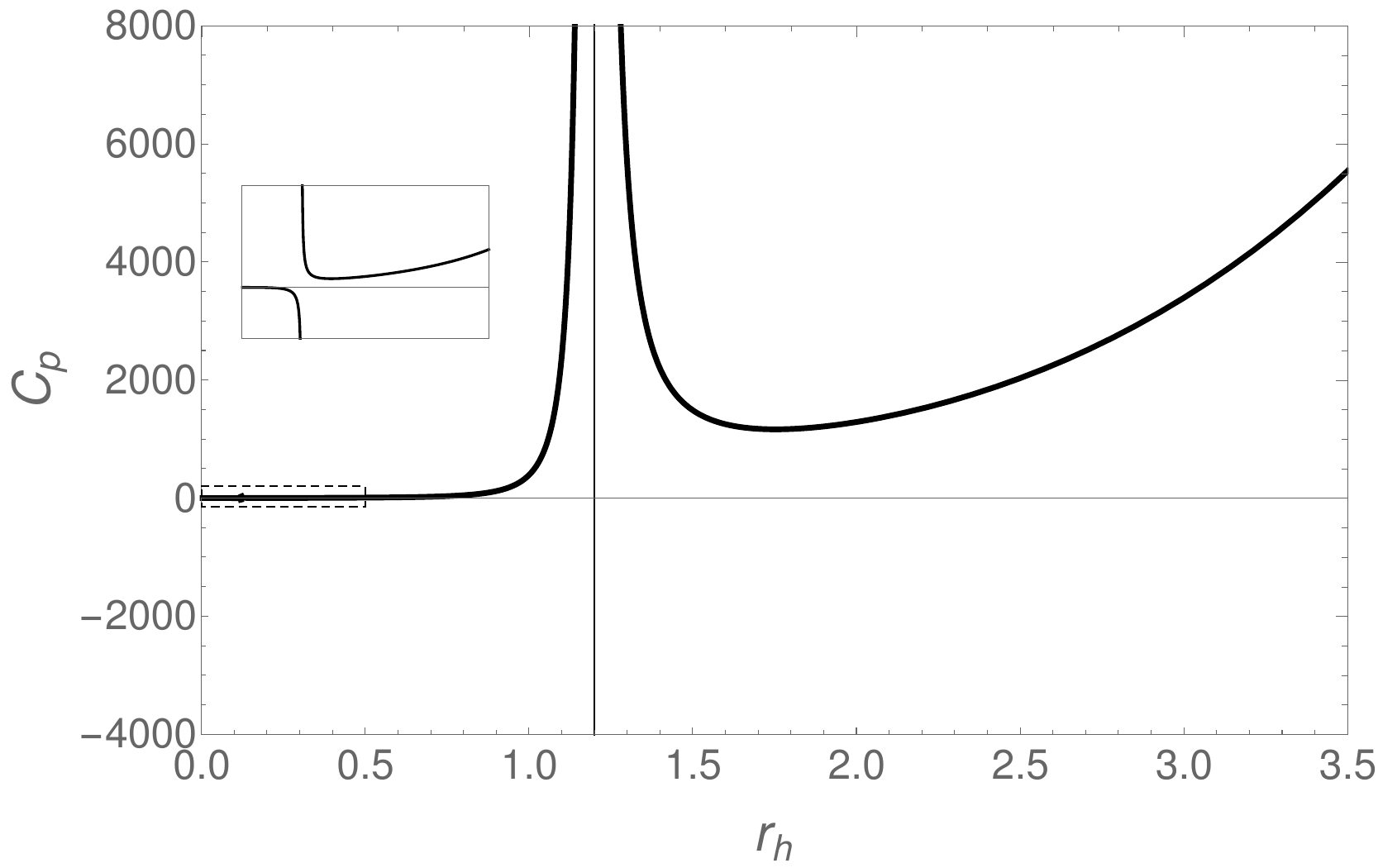}\\
  \includegraphics[width=0.4\linewidth]{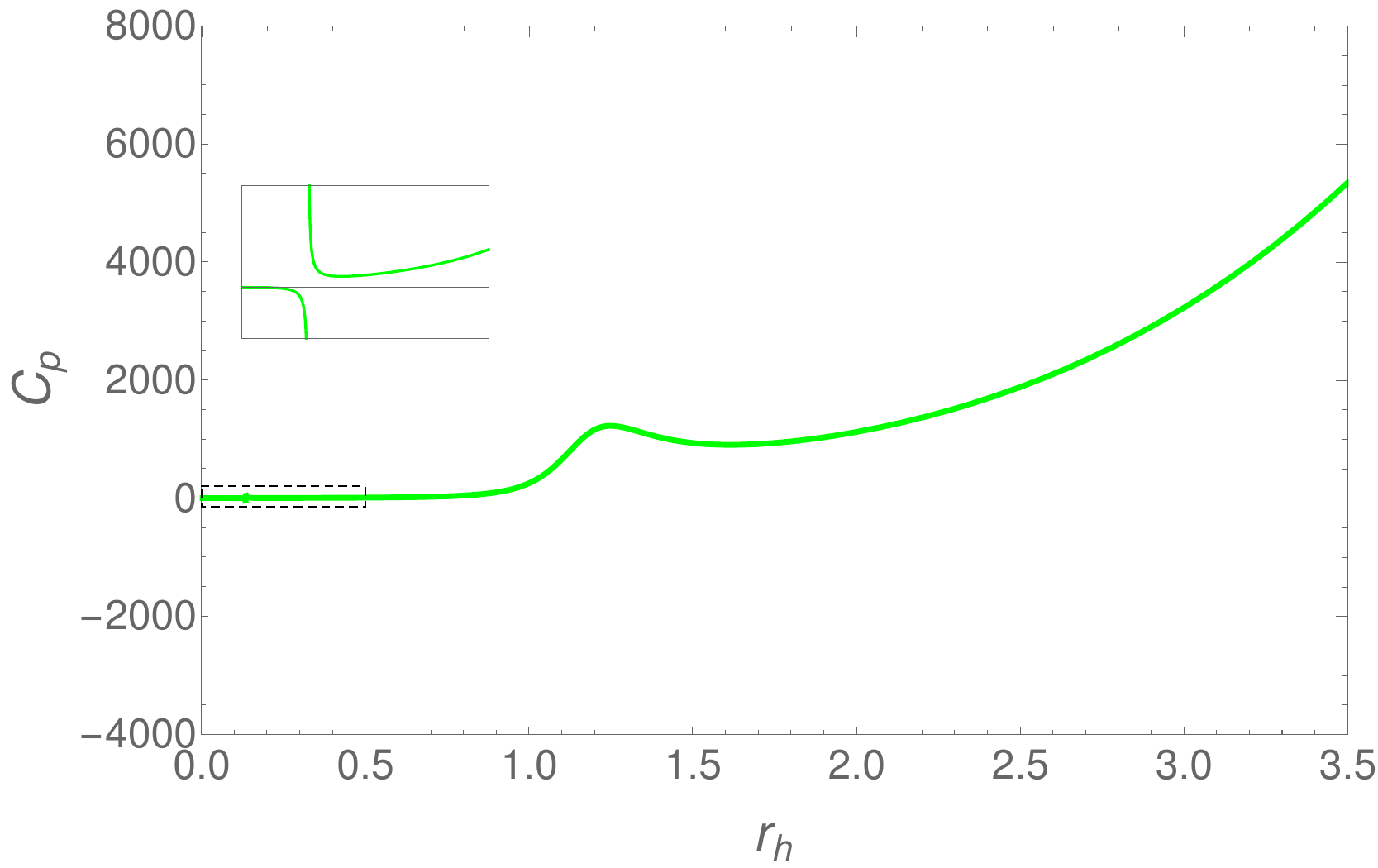}
  \caption{The heat capacity $C_P$ at the constant pressure as a function of the event horizon radius for $D=6$ in three cases: $P<P_c$ (top-left), $P=P_c$ (top-right), and $P>P_c$ (bottom). The small panels are the zoom of the dashed rectangles.}
  \label{fig:cp6}
\end{figure}
In Fig. \ref{fig:FD6}, we find that black holes are thermodynamically unstable at the regime of the relatively small $r_h$ because they correspond to the local maximum of the generalized free energy. This is also observed in Fig. \ref{fig:cp6} where these black holes have a negative heat capacity. The thermodynamically unstable phase is separated from the thermodynamically stable phase by a divergent line which indicates a second-order phase transition between these phases. 
\begin{figure}[ht]
  \includegraphics[scale=0.3]{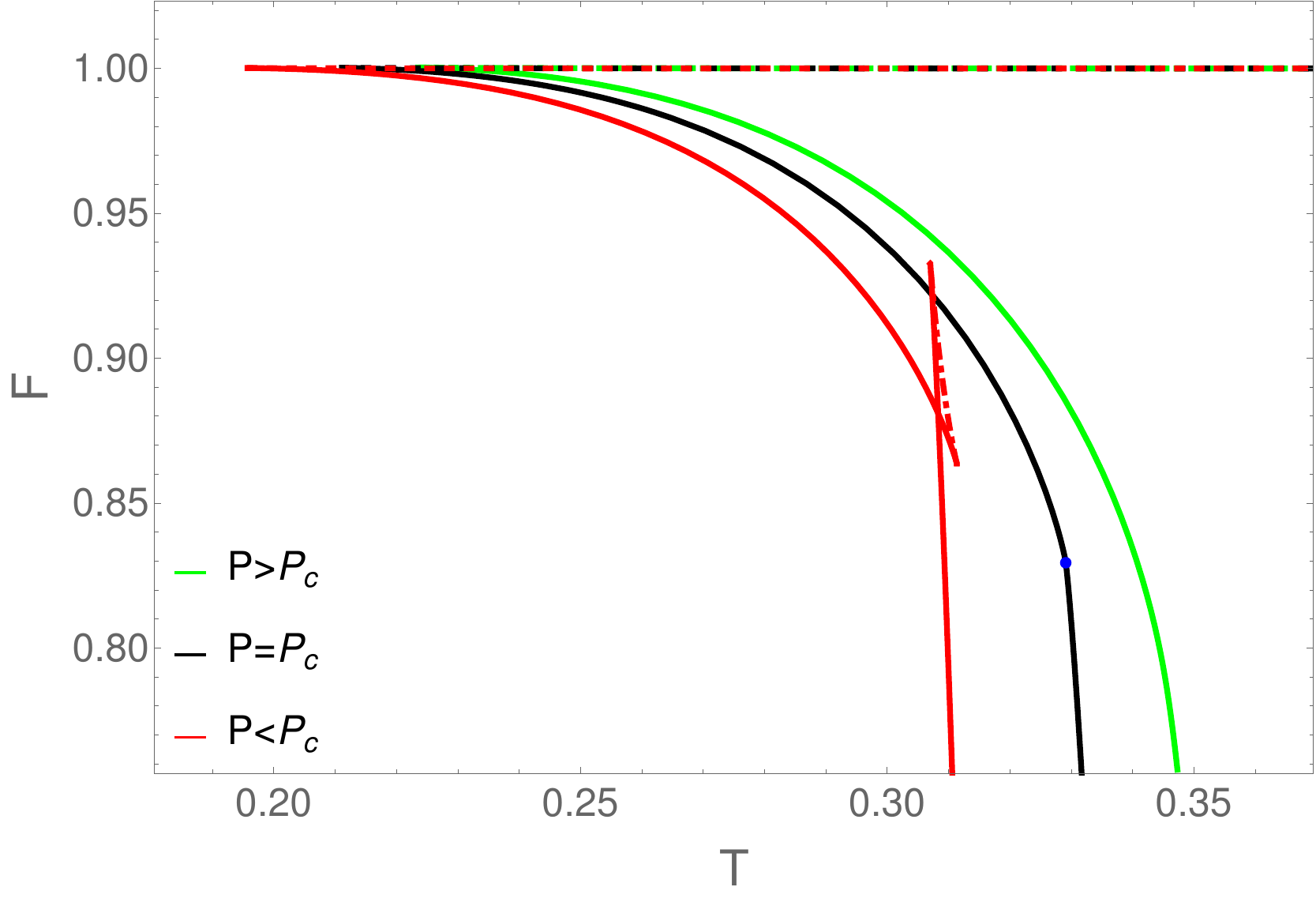}
  \caption{The behavior of the on-shell free energy in terms of the black hole temperature for $D=6$ in three cases: $P<P_{c}$ (red), $P=P_{c}$ (black), and $P>P_{c}$ (green). The dashed curves refer to the unstable phases which have a negative heat capacity. The blue dot is the critical point of the Van der Waals-like phase transition.}
  \label{fig:gibbs6}
\end{figure}
In Figs. \ref{fig:gibbs6}, we also observe the presence of these two thermodynamic phases and the corresponding phase transition with an arbitrary pressure. The thermodynamically unstable small black holes are represented by the dashed lines with the on-shell free energy close to one. This branch is connected to the branch of the thermodynamically stable black holes represented by the solid curves with the lower on-shell free energy. This connection leads to the multivalued behavior of the on-sell free energy indicating the second-order phase transition.

When the pressure is lower than the critical value $P_c$, there are two additional second-order phase transitions between the thermodynamically stable small(large) black holes and the thermodynamically unstable intermediate black hole. In addition, there is a first-order phase transition (or the Van der Waals-like phase transition) between the small black holes and large black holes. This reveals the presence of two divergent lines in the behavior of the heat capacity which separate the negative regime from the positive one. The two thermodynamic phases corresponding to the intermediate and large black holes are exhibited in the behavior of the generalized free energy $F$ with the occurrence of additional local extremal points which are the on-shell black hole states when $P<P_c$. These thermodynamic phases and the corresponding phase transitions are analogous to those in the second class $D=5$.

\section{The thermodynamic phases in the grand canonical ensemble}
\label{sec:GCE}

\begin{figure}[ht]
  \includegraphics[scale=0.3]{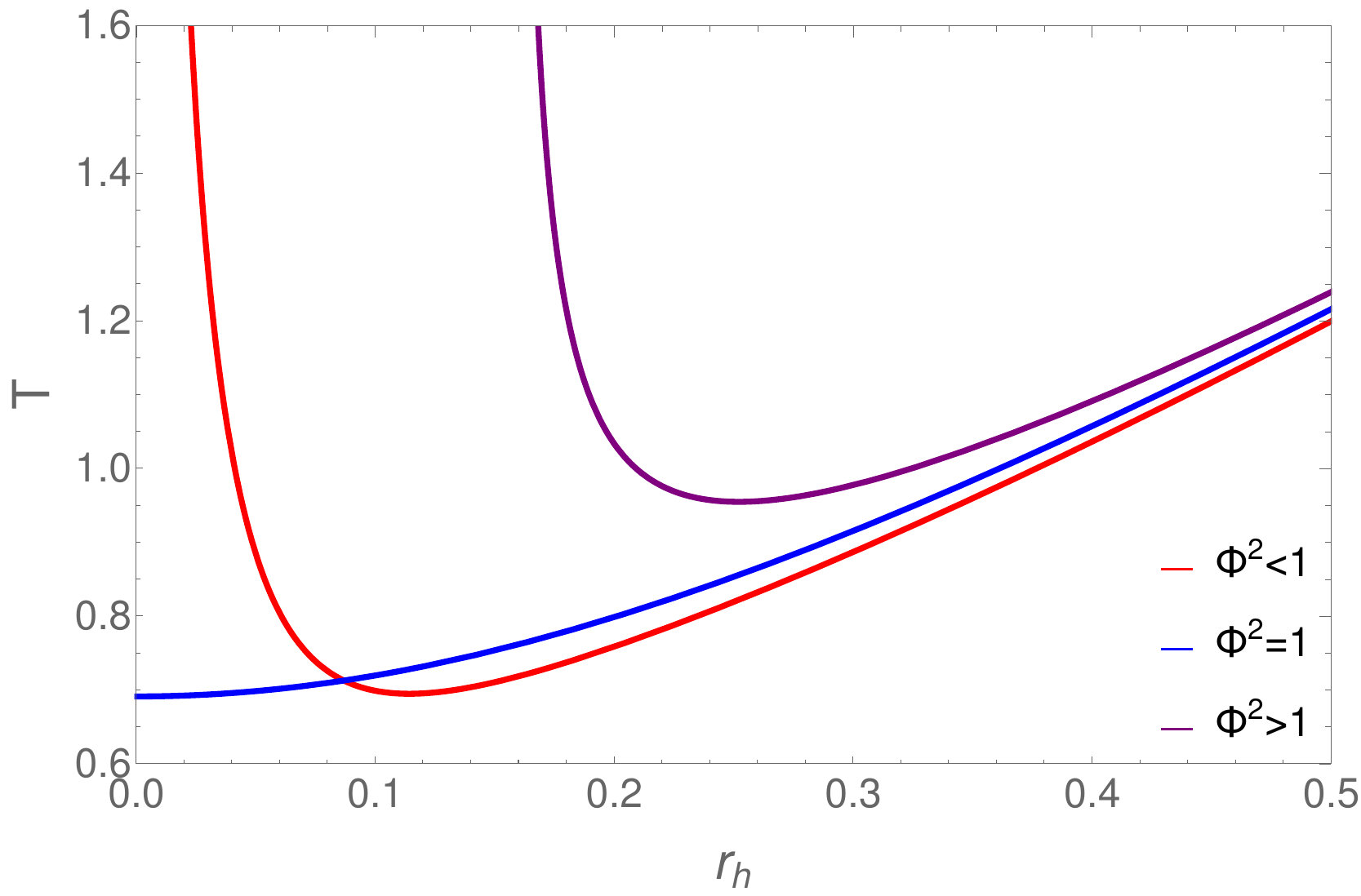}
  \caption{The Hawking temperature versus the event horizon radius for $D=4$ with various chemical potential values.}
  \label{fig:gce.T4}
\end{figure}

We study the thermodynamic phases and the phase transitions of black holes in the grand canonical ensemble where the chemical potential $\Phi$ is kept fixed. From Eqs. (\ref{eqn:potential}) and (\ref{ehrad:eq}), we express the parameters $q$ and $m$ in terms of $\Phi$ and $r_h$ as follows
\begin{eqnarray}
    q&=&\frac{r_{h}^{D-3}\Phi^2}{1+g^2r_{h}^2-\Phi^2},\label{paeq:gcen}\\
    m&=&\frac{r_{h}^{D-3} \left[\left(1+g^2 r_{h}^2\right)^2-\Phi ^2\right]}{1+g^2 r_{h}^2-\Phi ^2}.\label{paem:gcen}
\end{eqnarray}
Using these equations, we write the generalized Gibbs free energy in the grand canonical ensemble as follows
\begin{eqnarray}
\Omega&=&M-\tau S-\Phi Q\nonumber\\
&=&\frac{\omega_{D-2}}{16\pi}r^{D-3}_h\left[\frac{\Phi^2\left(\Phi^2-1\right)}{1+g^2r^2_h-\Phi^2}+\left(D-2\right)\left(1+g^2r^2_h\right)+\Phi^2-4\pi\tau r_h\sqrt{\frac{1+g^2r^2_h}{1+g^2r^2_h-\Phi^2}}\right].
\end{eqnarray}
By taking the first-order derivative of the generalized free energy in the event horizon radius to be zero, we can obtain the Hawking temperature which is rewritten in the grand canonical ensemble by using Eqs. (\ref{paeq:gcen}) and (\ref{paem:gcen}) as follows
\begin{eqnarray}
    T=\frac{D-3+(D-1)g^2r_{h}^2-(D-3)\Phi^2}{4\pi r_h}\sqrt{\frac{1+g^2r_{h}^2}{1+g^2r_{h}^{2}-\Phi^2}}.
\end{eqnarray}
From this expression, we find a constraint on the event horizon radius given by
\begin{eqnarray}
    r_{h}^{2}>\frac{\Phi^2-1}{g^2}.\label{eqn:cons}
\end{eqnarray}
This constraint implies that when $\Phi^2>1$, the event horizon radius $r_h$ should have a lower bound. The behavior of the Hawking temperature in terms of the event horizon radius crucially depends on the chemical potential $\Phi$, as depicted in Fig. \ref{fig:gce.T4}. 
For $\Phi^2\neq1$, the Hawking temperature increases to infinity as $r_h$ approaches either the lower bound (which is nonzero for $\Phi^2>1$) or the infinity. In this case, the Hawking temperature possesses a global minimum. However, in the case of $\Phi^2=1$, the Hawking temperature is a monotonously increasing function of $r_h$. When $r_h$ goes to zero, the Hawking temperature approaches a minimum value given by
\begin{eqnarray}
T_{\text{min}}=\sqrt{\frac{D-1}{(D-2)\pi}}.
\end{eqnarray}

Unlike the thermodynamics in the canonical ensemble which depends on the number of spacetime dimensions and the pressure, the thermodynamics in the grand canonical ensemble depends on the chemical potential. It is classified into the following classes: (i) $\Phi^2<1$; (ii) $\Phi^2>1$; and (iii) $\Phi^2=1$.
\begin{figure}[ht]
  \includegraphics[scale=0.33]{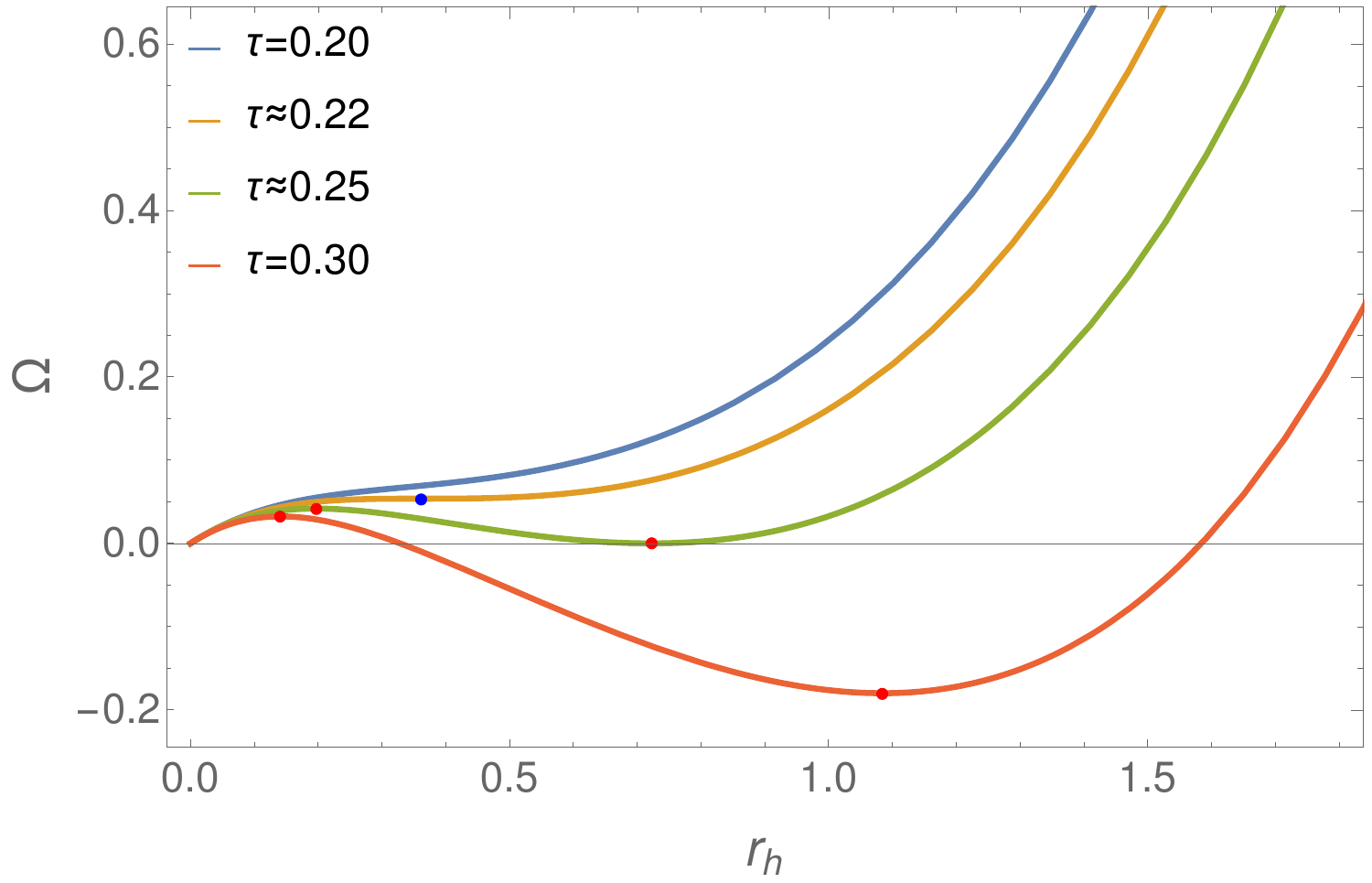}
  \includegraphics[scale=0.33]{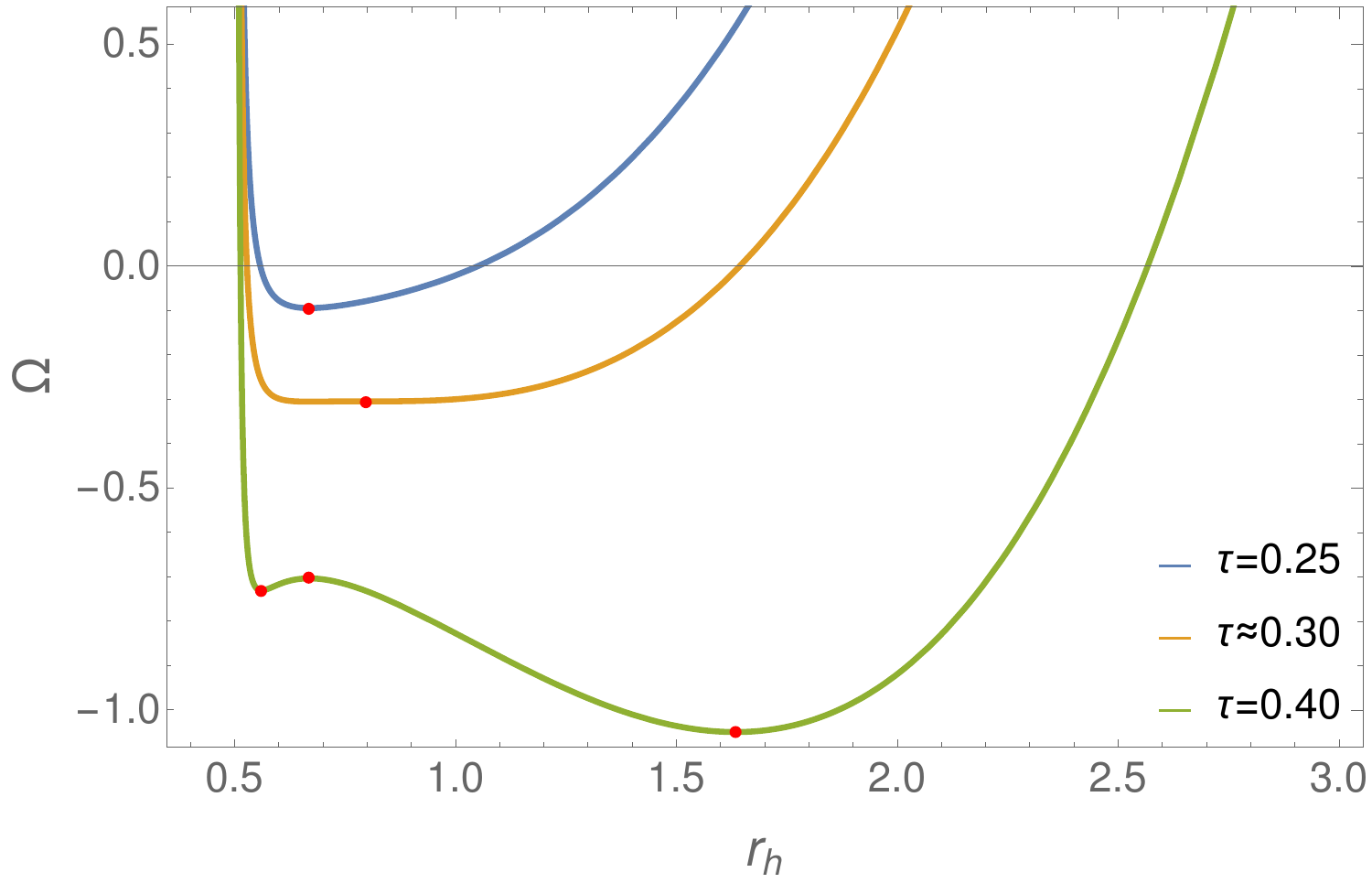}\\
  \includegraphics[scale=0.33]{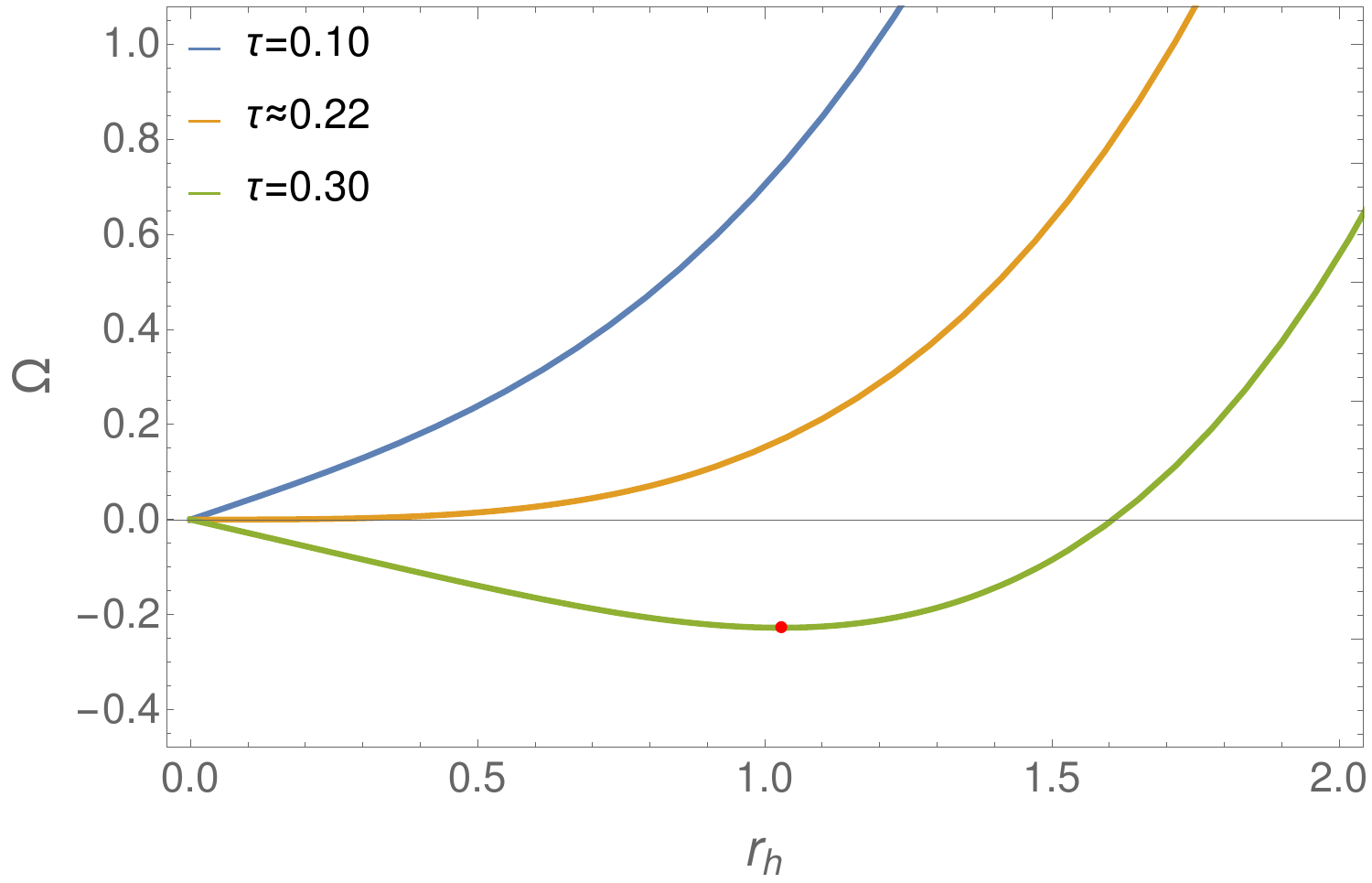}
  \caption{The generalized free energy $F$ in terms of the event horizon radius $r_h$ for various values of the ensemble temperature and the chemical potential with $D=4$ and $P=0.1$. The top, middle, and bottom panels correspond to $\Phi^2=0.9$, $1$, and $1.1$, respectively. The red points represent the on-shell black hole states corresponding to the black holes. The blue points represent the inflection point.}
  \label{fig:gce.FD4}
\end{figure}

\begin{figure}[ht]
  \includegraphics[scale=0.3]{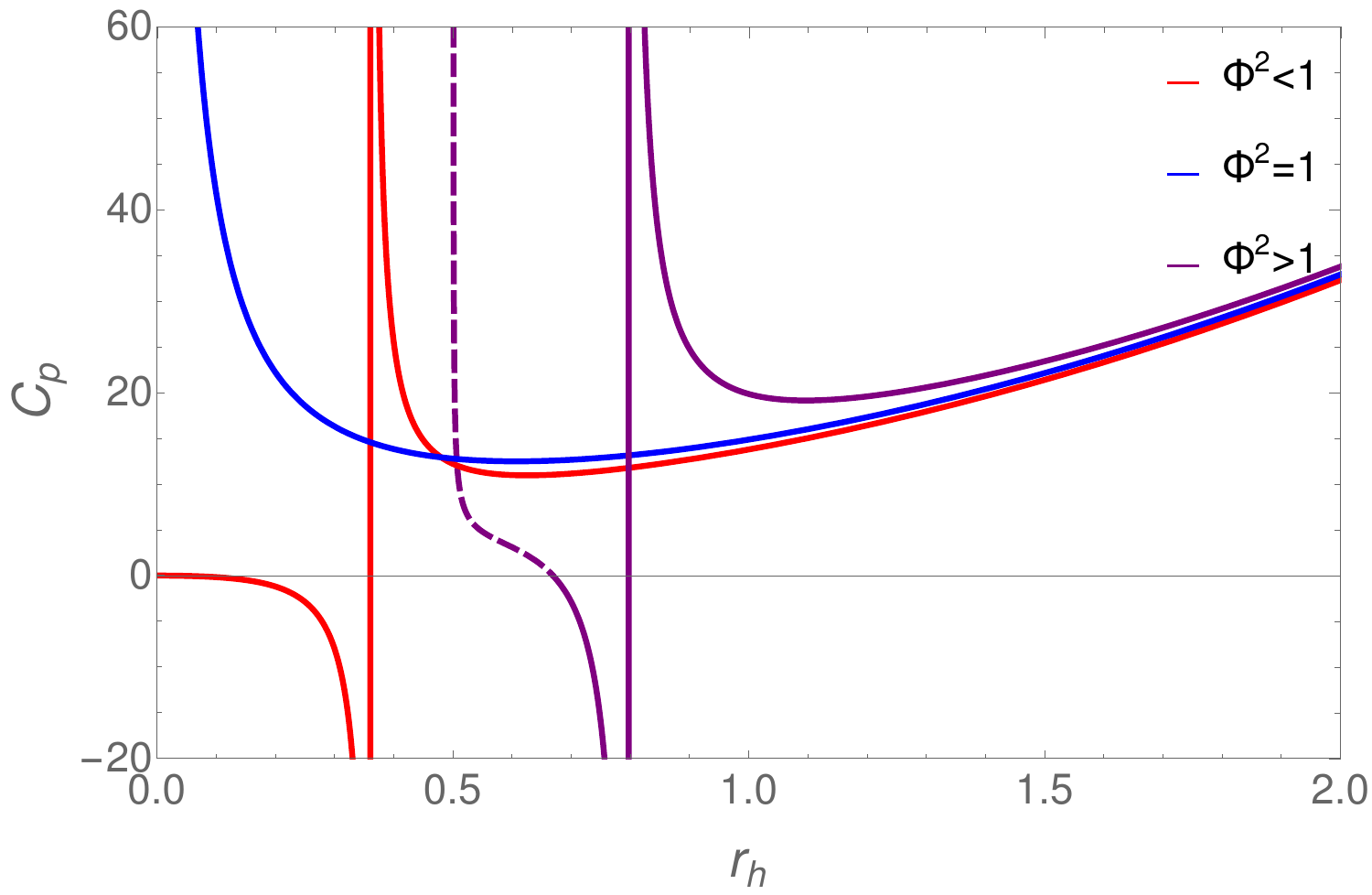}
  \caption{The heat capacity at the constant pressure $C_P$ versus the event horizon radius for various chemical potentials with $D=4$ and $P=0.1$.}
  \label{fig:gce.Cp4}
\end{figure}

\begin{figure}[ht]
  \includegraphics[scale=0.3]{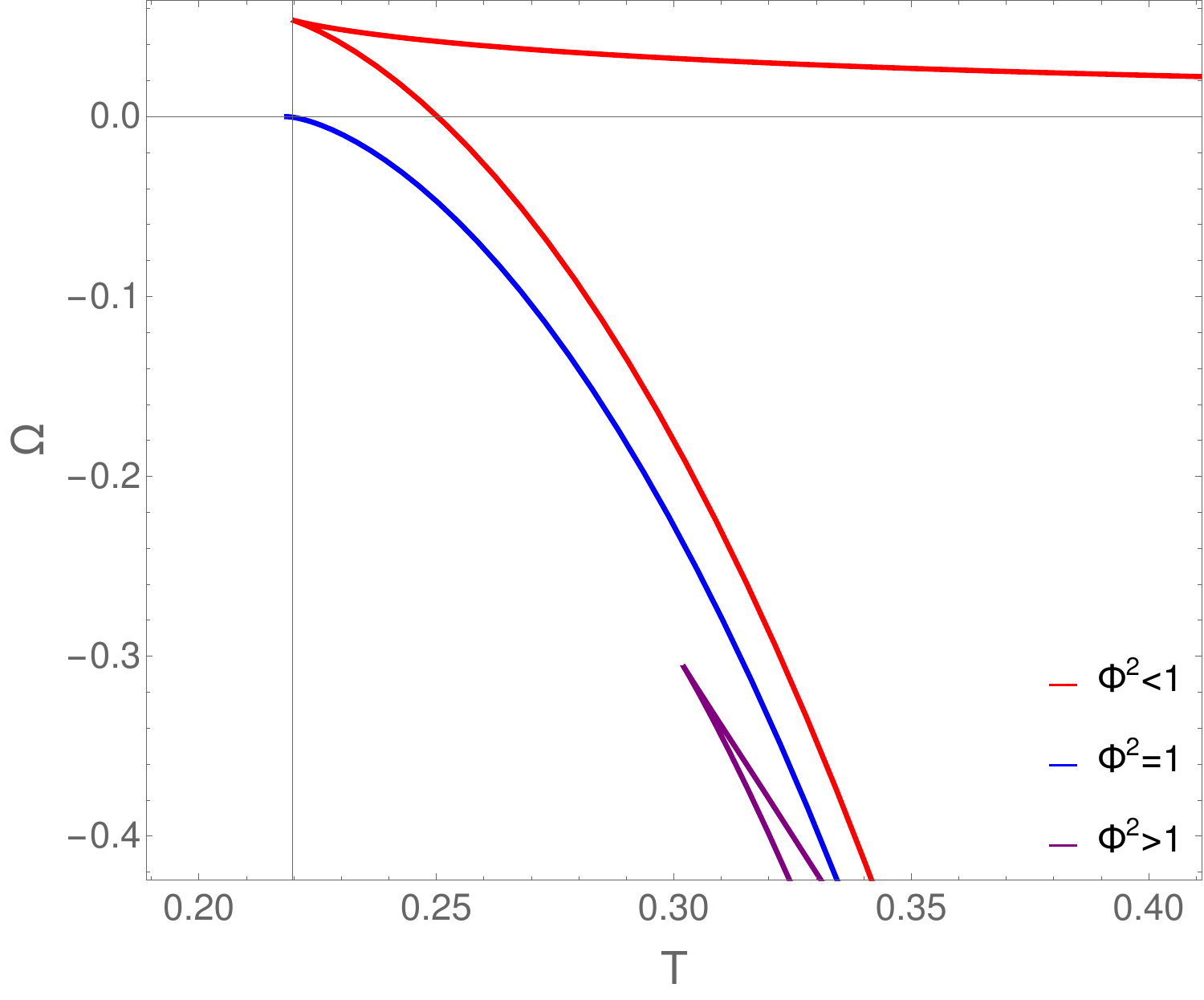}
  \caption{The on-shell free energy in terms of the event horizon radius for $D=4$ and $P=0.1$.}
  \label{fig:gce.G4}
\end{figure}

The thermodynamics of the $\Phi^2<1$ class is the same as that of the $D=4$ class in the canonical ensemble. When the ensemble temperature is above a minimum value, the generalized free energy $\Omega$ exhibits a local maximum and global minimum which indicate the on-shell black hole states, as depicted by the top panel in Fig. \ref{fig:gce.FD4}. The local maximum of the generalized free energy represents the thermodynamically unstable black holes which would decay into the thermodynamically stable black holes represented by the global minimum. This decay corresponds to a second-order phase transition which can also be pointed by studying the behavior of the heat capacity at the constant pressure $C_P$ and the on-shell free energy $G$ which are given by
\begin{eqnarray}
    C_P&=&\frac{2\pi r_{h}^{2}(1+8\pi r_{h}^{2}-\Phi^2)\left[(3+8\pi r_{h}^{2})^{2}-9(1+4\pi r_{h}^{2})\Phi^2\right]}{(8\pi r_{h}^{2}-1)(3+8\pi r_{2}^{2})^{2}+2\left[9-4\pi r_{h}^{2}(3+4\pi r_{h}^{2})\right]\Phi^2-9\Phi^4}\sqrt{\frac{3+8\pi r_{h}^{2}}{3+8\pi r_{h}^{2}-3\Phi^2}},\\
    \Omega&=&M-TS-\Phi Q\nonumber\\
    &=&-\frac{2\pi r_{h}^{3}}{3}+\frac{r_h(3+8\pi r_{h}^{2})}{4(3+8\pi r_{h}^{2}-3\Phi^2)}.
\end{eqnarray}    
As seen in Fig. \ref{fig:gce.Cp4}, there is a divergence in the behavior of $C_P$ for $\Phi^2<1$ (the red curve).
This divergence separates the heat capacity into two thermodynamic phases: the thermodynamically (un)stable phase with the (negative) positive $C_P$ corresponds to the (small) large black holes. Also, From in Fig. \ref{fig:gce.G4}, 
we find that the free energy of the small black holes (shown by the red line above the horizontal axis) is higher than that of the large black holes (shown by the remaining branch of the red curve). This means that the small black holes are less thermodynamic than the large black holes. In addition, the on-shell free energy shows multivalued behavior, which indicates the phase transition between the small and large black holes. Furthermore, we find the HP phase transition between the thermal AdS space and the large black holes when the global minimum of the generalized free energy lies on the line $\Omega=0$. The HP temperature can be determined by the intersection between the lower branch of the red curve (given in Fig. \ref{fig:gce.G4}) with the line $\Omega=0$.

For the $\Phi^2>1$ class, when the ensemble temperature is below a critical value which is about $0.3$ for $D=4$ and $P=0.1$, the generalized free energy has a global minimum corresponding to the on-shell black hole states which are thermodynamically stable or have the positive heat capacity. However, when the ensemble temperature is above the critical value, we observe from the middle panel of Fig. \ref{fig:gce.FD4} that there are one local maximum, one local minimum, and one global minimum. The local minimum with the small $r_h$ represents the thermodynamic phase with a positive heat capacity as shown by the dashed purple curve in Fig. \ref{fig:gce.Cp4}. The local maximum represents the thermodynamically unstable black holes with a negative heat capacity as shown by the purple curve below the horizontal axis in Fig. \ref{fig:gce.Cp4}. The global minimum represents the thermodynamically stable black holes that have a positive heat capacity as shown by the purple curve above the horizontal axis in Fig. \ref{fig:gce.Cp4}. The unstable black holes would decay into the stable black holes corresponding to the local minimum and global minimum via some phase transition. However, it should be noted that the transition between the on-shell black hole states corresponding to the local minimum and maximum is a crossover, not a phase transition because the thermodynamic quantities are continuous.

For the $\Phi^2=1$ class, from the bottom panel of Fig. \ref{fig:gce.FD4}, we see that when the ensemble temperature is above a minimum value, the generalized free energy possesses a unique extremal point, implying that there is a unique thermodynamic phase. This phase has a positive heat capacity, as shown by the blue curve in Fig. \ref{fig:gce.Cp4}, and hence it is thermodynamically stable. This is consistent with the fact that the extremal point of the generalized free energy is a global minimum. On the contrary, when the ensemble temperature is below this minimum value, the thermodynamic system exits at the thermal AdS space. In this sense, this minimum value is nothing but the HP phase transition temperature which is about $0.22$ for $D=4$ and $P=0.1$. We can determine this temperature from the intersection of the blue curve (given in Fig. \ref{fig:gce.G4}) with the horizontal axis.

\section{\label{sec:conclu} Conclusion}

Gauging the KK vector field leads to a pseudo-supersymmetrization of the $D$-dimensional KK theory (derived from the KK reduction of $(D+1)$-dimensional pure Einstein gravity on a circle $S^1$) where the pseudo-gravitino and pseudo-dilatino are charged under U(1)$_{\text{KK}}$. As a result, this gauge generates a potential for the dilaton field. The charged dilaton $\text{AdS}$ black holes in this effective field theory have been found in Ref. \cite{Liu2012} which are asymptotic to the AdS$_D$ geometry where the curvature radius of asymptotic geometry is determined in terms of the KK gauge coupling. Interestingly, these black holes can be obtained from the dimensional reduction of supergravity on the compact internal manifolds. This allows us to restrict the number of spacetime dimensions as $4\leq D\leq 7$.

In the present work, we study the thermodynamic phases and the phase transitions of charged dilaton $\text{AdS}$ black holes in the context of the generalized (off-shell) free energy which is an extension of the on-shell free energy. When the ensemble temperature is below a minimal value, the generalized (off-shell) free energy has a global minimum located at the zero event horizon radius (without having any extremum point) corresponding to the free energy of the thermal AdS space. This implies that the thermodynamic system exits at the phase of the thermal AdS space. However, the generalized free energy would exhibit local maximum and local/global minimum when increasing the ensemble temperature above the minimum value. These extremal points correspond to the on-shell states which satisfy the equations of motion. The local maximum points represent the thermodynamically unstable black holes with negative heat capacity. Hence, they would decay into the thermodynamically stable black holes (with a negative heat capacity) represented by the local/global minimum points. These decays are performed through either the phase transition or the crossover. This is confirmed by studying the divergence of the heat capacity at the constant pressure and the multivalued behavior of the on-shell free energy. Investigating the generalized (off-shell) free energy, we also find the HP phase transition between the thermal AdS space and black holes, and the phase transition between the small black holes (represented by local minimum) and the large black holes (represented by global minimum) analogous to the Van der Waals phase transition between the liquid and gas.

In particular, we have classified the thermodynamics of charged dilaton $\text{AdS}$ black holes in the gauged KK theory in both the canonical ensemble and grand canonical ensemble. The thermodynamic phases and phase transitions of black holes in the canonical ensemble are dependent on the number of spacetime dimensions and the pressure, consequently, the thermodynamic behavior of the black hole can be classified into three different classes corresponding to $D=4$, $D=5$, and $D=6,7$. In the grand canonical ensemble, the thermodynamics of black holes depends only on the chemical potential $\Phi$ and can be classified into three different classes corresponding to $\Phi<1$, $\Phi>1$, and $\Phi=1$.

\section*{Acknowledgments}
This research is funded by Phenikaa University under grant number PU2022-1-A-16.

\end{document}